\documentstyle[preprint,aps,pre]{revtex}
\tighten
\begin{document}
\input epsf
\begin{figure}
\end{figure}
\begin{table}
\end{table}
\newpage
\setcounter{page}{1}
\preprint{WIS-95-ED}
\draft
\title{ Lack of Self-Averaging in Critical Disordered Systems }
\author{Shai Wiseman and Eytan Domany}
\address{Department of Physics of Complex Systems, Weizmann Institute of
science, \\          Rehovot 76100 Israel }
\date{June 18, 1995}
\newcommand{\av}[1]{\langle #1 \rangle}
\maketitle
\begin{abstract}
We consider the sample to sample fluctuations that occur in the value of a
thermodynamic quantity $P$ in an ensemble of finite systems with quenched
disorder, at equilibrium. The variance of $P$, $V_{P}$, which characterizes
these fluctuations is calculated as a function of the systems' linear size
$l$, focusing on the behavior at the critical point. The specific model
considered is the bond-disordered Ashkin-Teller model on a square lattice.
Using Monte Carlo simulations, several bond-disordered Ashkin-Teller
 models were examined, including the bond-disordered Ising model and the
bond-disordered four-state Potts model. It was found that far from criticality
 all thermodynamic quantities which were examined (energy, magnetization,
specific heat, susceptibility) are strongly self averaging, that is
$V_{P}\sim l^{-d}$ (where $d=2$ is the dimension). At criticality though, the
results indicate that the magnetization $M$ and the susceptibility $\chi$ are
 non self averaging, i.e. $\frac{V_{\chi}}{\chi^{2}},\frac{V_{M}}{M^{2}}\not
\rightarrow 0$. The energy $E$ at criticality is clearly weakly self
averaging, that is $V_{E}\sim l^{-y_{v}}$ with $0<y_{v}<d$. Less conclusively,
and possibly only as a transient behavior, the specific heat too is found to
be weakly self averaging. A phenomenological theory of  finite size scaling
for disordered systems is developed, based on physical considerations similar
to those leading to the Harris criterion. Its main prediction is that when the
specific heat exponent $\alpha<0$ ($\alpha$ of the disordered model) then,
for a quantity $P$ which scales as $l^{\rho}$ at criticality, its variance
$V_{P}$ will scale asymptotically as $l^{2\rho+\frac{\alpha}{\nu}}$. The
theory is not applicable in the asymptotic limit ($l\rightarrow \infty$) to
the bond-disordered Ashkin-Teller model where $\frac{\alpha}{\nu}=0_{+}$.
Nonetheless in the accessible range of lattice sizes we found very good
agreement between the theory and the data for $V_{\chi}$ and $V_{E}$.
The theory may also be compatible with the data for the variance of
the magnetization $V_{M}$ and the variance of the specific heat $V_{C}$, but
evidence for this is less convincing.

\end{abstract}

\narrowtext
\section{introduction}

 How is the critical behavior affected by the introduction of disorder
(usually dilution or bond--randomness) into a model? This question has been
extensively studied\cite{Stinch} experimentally,
analytically\cite{Shalaev94} and
numerically\cite{Sel Sing} for quite some time now. Many studies concentrate
on finding out to which universality class certain disordered models
belong, e.g. calculating critical exponents. In
this work we consider a different aspect of the same question.
The measurement of any {\em density } of an extensive thermodynamic property
$P$ (e.g. $P=E,M,C_{h}$ or $\chi$) in a disordered
system may hypothetically be done in the following way. An ensemble of
macroscopic disordered samples of size $l$ is prepared; denote by $x$ a
sample with a particular random realization of
the quenched disorder. Now in each sample $x$, $P_{x}(t)$
 is measured over a long time interval, and $\overline{P_{x}}$, the average
over time $t$ is calculated.
Close to the critical point the measurement of $\overline{P_{x}}$ will
require long times due to large thermal fluctuations which will occur.
In addition, since in every sample a different configuration of the quenched
disorder is present, a different value for $\overline{P_{x}}$ will be
measured.
Next, the average of $\overline{P_{x}}$ over the ensemble
$[\overline{P_{x}}]$  ($[\ldots]$ stands for an ensemble average over the
different samples) is calculated and so is its variance \begin{equation}
 V_{P}=[ \{ \overline{P_{x}}-[\overline{P_{x}}] \}^{2} ]
\;. \label{eq: V_p} \end{equation}
 Assume that the
time interval of the measurement was long enough so that thermal
fluctuations in $P_{x}(t)$ were averaged out perfectly and
$\overline{P_{x}}$ may be considered to be exact. The question then rises:
How will the variance $V_{P}$ change as the critical temperature is
approached or as the correlation length $\xi$\cite{accurate} is increased?
This question, which concerns the way in which disorder affects the behavior
of systems near their critical point, is approached in this work
using the framework of finite size scaling.

A common practice in Monte Carlo (MC) simulations is to examine the critical
behavior
by simulating a system at its critical temperature $T_{c}^{\infty}$ and
changing the lattice
size $l$. According to the theory of finite size scaling\cite{Barber} the
lattice size $l$ sets the
scale of the correlation length in such a finite system. Thus the dependence
of $P$ on $\xi$ in an infinite system close to criticality is substituted by
dependence on $l$ in a finite system at criticality. When a disordered
system is considered then many samples need to be simulated in order to
obtain estimates of $P$ which are averaged over the disorder. In this case,
the question, which is the main theme of this work, would
be: how does the effect of disorder on the sample to sample fluctuations  in
$\overline{P_{x}}$  change, as
the lattice size $l$ is increased at
the critical temperature? Or how does $V_{P}$ scale with $l$? This
question is not only of theoretical interest in its own right, but also of
practical interest for MC studies of critical disordered systems. If the
relative variance $V_{P}/[P]^2$
decreases with increasing $l$ then the number of samples needed to obtain
[P] to a given accuracy goes down with increasing $l$. If, on the
other hand, $V_{P}/[P]^2$ is independent of $l$, then the number of samples
which need to be simulated is independent of $l$ and the total amount of work
rises very strongly with $l$.

The issue which we study in this work should not be confused with two
closely related issues. The first is usually referred to as the property of
self-averaging of additive (extensive) quantities in disordered
systems\cite{Rev:Mod}.
Consider again the ensemble of macroscopic disordered samples of size $l$.
The question is then whether
\begin{equation}
 V_{P}/[P]^2 \rightarrow 0 \;\;\;\; {\text as } \; l\rightarrow
\infty
\;. \label{eq: Self} \end{equation}
If so, then the measurement of $\overline{P_{x}}$ in one very large sample
$x$ which occurs with reasonable probability will provide a good estimate of
the ensemble average.
This is very important for the comparison of theoretical work, where the
configurational average is taken,  with experiments, where only a
large single sample is examined.
As first argued by Brout\cite{Brout}, we may divide the sample $x$ into $n$
 large subsamples (much larger than the correlation length $\xi$). If
we assume that the coupling between neighboring subsystems is negligible,
then the value of any density of an extensive quantity over the whole sample
is equal to the average of the (independent) values of this quantity over
the subsamples.
Provided the probability distribution of the $P$'s of the subsamples has a
finite variance, then according to the Central Limit theorem the value of
$\overline{P_{x}}$ is distributed with a Gaussian probability
distribution around its mean $[\overline{P_{x}}]$. The square of the width
of the
Gaussian, $V_{P}$, is proportional to $\frac{1}{n}\sim l^{-d}$.
In this case (\ref{eq: Self}) is fulfilled, and $P$ is called self-averaging.

 The quantities which are studied here
are all densities of extensive self averaging quantities (far from
criticality). Nonetheless,
note that our question, as it was formulated for macroscopic samples ($l\gg
\xi$), concerned the dependence of $V$ on the correlation length $\xi$ and
not on
the sample size $l$. On the other hand, as we will examine finite samples of
size
$l$ at criticality where $\xi\sim l$, the Brout argument does not hold,
since the average of $P$ over neighboring subsamples may not be considered as
independent. Thus at criticality there is no reason to expect that
$V_{P}\sim l^{-d}$.
An example for a phase transition, where sample to sample fluctuations
result
in non self averaging of certain quantities, is the percolation transition.
It has been shown\cite{percolation} that the resistive susceptibility and
the conductivity are non self averaging at the percolation threshold.

A second related issue is that of self-averaging in homogeneous systems.
This question concerns the thermal fluctuations in the value of a density
$P$ in a homogeneous system of size $l$. Define the thermal variance
as $\sigma^{2}_{T}=\av{ (P -\av{P})^2 }$, where $\av{\ldots }$ denotes
thermal
or time averaging. The following notions (slightly modified) have been
introduced by Milchev Binder and Heermann\cite{BH:book,self:av}:
If $\sigma^{2}_{T}/\av{P}^2\rightarrow 0$ as $l \rightarrow \infty$ then $P$
is self averaging otherwise it is said to exhibit lack of self averaging. If
\begin{equation}
\sigma^{2}_{T}/\av{P}^2\sim l^{-d}
 \label{eq: strongly} \end{equation}
 then $P$ is strongly self averaging. If
\begin{equation}
\sigma^{2}_{T}/\av{P}^2\sim l^{-x_{1}}\;\;\;\;\; {\mbox and }\;\;\;\;\; 0<
x_{1}\leq d  \label{eq: weakly} \end{equation}
 then $P$ is weakly self averaging. When $l\gg \xi$ it was
found\cite{BH:book,self:av} that averages of simple densities such as  $E,M$
are strongly self averaging while quantities obtained from the fluctuations
of these densities such as the specific heat $C$ and susceptibility $\chi$
are non self averaging. At criticality the singular part of the energy $E$
is weakly self averaging while $C$, $M$ and
$\chi$ exhibit lack of self averaging.
For example $\av{M}^2\sim l^{-2\beta/\nu}$ and $\sigma^{2}_{T,M}
\sim \chi/l^d \sim l^{\gamma/\nu-d}=l^{-2\beta/\nu}$, so that $M$ is non
self averaging.

The issues of self averaging in disordered systems and homogeneous systems
concern
the asymptotic behaviour of the fluctuations due to disorder and the thermal
 fluctuations respectively as the system size is increased.
While self averaging in homogeneous systems at criticality has been
addressed previously\cite{BH:book,self:av}, this study involves the question
of self averaging in disordered systems at criticality. With the increase
in the available computational power, a numerical investigation of the
sample to sample fluctuations of thermodynamic quantities is nowadays
feasible (whereas previously only calculation of the ensemble
average, which is less demanding computationally, was feasible).


The particular model which is used here to study the question of the self
averaging of fluctuations due to disorder at criticality is the
bond-disordered Ashkin-Teller model on a square lattice. Actually this work
is based on further analysis of results which were obtained in a previous
 MC study\cite{RBAT} which aimed to determine the universality class of
 the model. The random-bond Ashkin-Teller model is particularly
suitable for studying the effects of disorder on critical behavior.
 This is because the pure model possesses a
line of critical points along which critical exponents vary continuously.
 In particular, the
scaling exponent corresponding to randomness $\phi=(\alpha/\nu)_{pure}$
 varies continuously and is positive. Thus, according to the Harris
 criterion\cite{Harris}, randomness is a relevant operator of varying
strength,
and the critical behaviour of the disordered model was indeed found to
differ from that of the pure system.
Our conclusion in the present work is that the effective exponent ratio
$\alpha/\nu$
of a disordered model plays a central role in determining the self averaging
of the fluctuations due to disorder at criticality. For the susceptibility,
for instance, our results agree very well with a finite size scaling theory
which we develop, according to which the relative variance of the
susceptibility, $V_{\chi}/[\chi]^2$, scales as $l^{\alpha/\nu}$ at the
critical temperature.
This implies lack of self averaging when $\alpha=0$ (as is found for the
random bond Ashkin-Teller model) and only weak self averaging for negative
$\alpha$. Our theory is successful also in describing, for models with
weak disorder, the effect of crossover on the variance.  \\
Our finite size scaling theory is very similar to the physical
arguments that lead to the Harris criterion\cite{Harris}, which was derived
near the pure system fixed point.
The difference is that we are assuming that similar considerations
are valid near the disordered fixed point as well.

This work is organized as follows. In section 1 we define the random bond
Ashkin-Teller model (RBAT) and summarize its critical properties as found
in a previous study\cite{RBAT}.
In sec. \ref{sec: vardef} we define various variances
of thermodynamic quantities in disordered systems and explain their meaning.
We explain how the `sample to sample variance' can be estimated from MC
results.
In section 3 we display our results for several bond disordered
Ashkin--Teller models, including the four-state Potts and Ising models.
We have measured the `sample to sample variance' at criticality for
different lattice sizes and also for different degrees of disorder.
We discuss some qualitative features of these results, such as the apparent
lack of self averaging and the dependence on the amount of disorder and on
the specific heat exponent $\alpha$. In section 4 we develop a
phenomenological finite size scaling theory for the `sample to sample
variance'. In section 5 we compare the predictions of the theory with the
numerical results. We find good agreement in the case of the
susceptibility and the energy, while the agreement in the case of the
specific heat and magnetization is more questionable.
\section{The random bond Ashkin-Teller model }
\label{sec:RBAT}
The model we study is the {\em Random-Bond} AT model (RBAT) on a
square lattice. On every site of the lattice two Ising spin variables,
$\sigma_i$ and $\tau_i$, are placed. Denoting by $<ij>$ a pair of nearest
neighbor sites, the Hamiltonian is given by
\begin{equation}
{\cal H}=-\sum_{<i,j>}[K_{i,j}\sigma_{i}\sigma_{j}+K_{i,j}\tau_{i}\tau_{j}+
\Lambda_{i,j}\sigma_{i}\tau_{i}\sigma_{j}\tau_{j}] \;.
   \label{eq:rbAT}  \end{equation}
 The positive coupling constants $K_{i,j}$ and $\Lambda_{i,j}$
 are chosen according to
   \begin{equation}
(K_{i,j},\Lambda_{i,j})=\left\{ \begin{array}{ll}
 \,(K^{1},\Lambda^{1}) & \mbox{with probability } \frac{1}{2} \\
 \,(K^{2},\Lambda^{2}) & \mbox{with probability } \frac{1}{2} \end{array}
\right. \label{eq:Ran} \;.   \end{equation}

The homogeneous model\cite{AT:1} [ $(K^{1},\Lambda^{1})=
(K^{2},\Lambda^{2})$
] possesses a line of critical points, along which critical exponents vary
continuously. This critical line interpolates between the Ising and four
state Potts models. Even though the scaling exponent corresponding to
randomness, $\phi=(\alpha/\nu)$, also varies continuously along this
line, it takes positive values, ($1\geq\phi \geq 0$), so that randomness is
relevant. Indeed the critical behaviour of the disordered model was found
to be different from that of the pure one\cite{RBAT}.
 In \onlinecite{RBAT} A duality
transformation was used to locate a critical plane of the disordered model;
The random model is critical when $(K^{2},\Lambda^{2})$ are the dual
couplings of $(K^{1},\Lambda^{1})$ \ \cite{Fisch78,Kinzel81,Wu74,Domany79}.
This critical plane corresponds to the line of critical points of the pure
model, along which critical exponents vary continuously.
A finite size scaling study was performed for several  critical models,
extrapolating
between the critical bond-disordered Ising and four state Potts models.
The critical behaviour of each disordered model was compared with the
critical behaviour of an anisotropic Ashkin-Teller model which was used as a
reference pure model\cite{pure}.
Whereas we found
no essential change induced by randomness in the order parameters' critical
exponents, the divergence of the specific heat $C$ did
change dramatically. Our results favor a logarithmic type divergence at
$T_{c}$, $C\sim \log l$  for the entire critical manifold of the random bond
Ashkin-Teller model, including the four
state Potts model, but excluding the random bond Ising model, for which
$C\sim \log \log l$ was obtained.

Here we give some of the details of the simulations and our main numerical
results for the critical behavior. These are necessary for understanding
and analyzing our variance results. All the results listed here were
presented in detail in \onlinecite{RBAT}; some essential points are reviewed
here for completeness sake.

Two series of critical RBAT models were studied in order to monitor two
effects. The first series of measurements were performed at five
models (or points in the couplings space),
$\{(K^{1},\Lambda^{1}),(K^{2},\Lambda^{2})\}_{i}\;i=0\ldots4$, which we
label as $C_{i},\;i=0\ldots4$. These were chosen so as to interpolate between
$C_{0}$, which is a random-bond Ising critical point (
$\Lambda^{1}=\Lambda^{2}=0$),
and $C_{4}$, which is a random-bond four-state Potts critical point
($\Lambda^{1}=K^{1}, \Lambda^{2}=K^{2}$).
The points $C_{i}$ interpolate in a similar manner to the way in which the
critical line of the pure AT connects the pure Ising critical point with the
pure four-state Potts critical point.
The extent of deviation from pure behavior is determined by the difference
between the two sets of couplings. For the series $C_{i},\;i=0\ldots4$
the ratio of $\frac{1}{10}$ was chosen, i.e.
\begin{equation}
 K^{2} \approx \frac{1}{10} K^{1}
\;,\label{eq:ratio1} \end{equation}
so that randomness will be pronounced\cite{Wang1,r=.5}.

Two additional measurement points (or models) were intended to monitor the
effect
of varying the amount of randomness on the critical behaviour.
The points $A_{2}$, $B_{2}$, $C_{2}$ represent three RBAT models with
coupling ratios $\frac{\Lambda^{1}}{ K^{1} } \approx \frac{1}{2}$ and
$\frac{ K^{2} }{ K^{1} } \approx \frac{1}{2},\;\frac{1}{4},\;\frac{1}{10}$
respectively. Thus the model $A_{2}$ possesses the lowest degree of
randomness, while the model $C_{2}$ possesses the highest degree of
randomness.
The usual definitions for energy $E$, specific heat $C$, magnetization
$M$\cite{magnetization},
susceptibility $\chi$, polarization $p=\av{\sigma\tau}$, and susceptibility
of the polarization $\chi^{(p)}$ were used. Since the specific heat seems
to play a dominant role in the behavior of the variance, we elaborate on
the specific heat results, and even reproduce one graph.
For the specific heat we found excellent agreement with the finite
size scaling form
 \begin{equation}
C=a_{0}+ b_{0} \ln [1+ c_{0}(l^{(\alpha/\nu)_{\text{pure}}}-1 )]\;,
\label{eq:C fss cross} \end{equation}
where $(\alpha/\nu)_{\text{pure}}$ is the critical exponent ratio of the
corresponding Anisotropic (pure\cite{pure}) model.
Eq. (\ref{eq:C fss cross}) reproduces expected scaling forms in various
limits as we now show. The constant $c_{0}$ can be expressed as
\begin{equation}
c_{0}= (l_{c}^{(\alpha/\nu)_{\text{pure}}}-1 )^{-1}
\;,  \label{eq:C lc} \end{equation}
where $l_{c}$ is a crossover length,
 at which crossover from the pure model's power law behaviour to the random
logarithmic behaviour occurs. Thus for $ l \ll l_{c}$ eq.
(\ref{eq:C fss cross}) reduces to the pure model behavior,
\begin{equation}
 C= a_{0}+b_{1} l^{(\alpha/\nu)_{\text{pure}}}  \;,
\label{eq: C(l)ani} \end{equation}
 while for $ l \gg l_{c}$ and $ l^{(\alpha/\nu)_{\text{pure}}}\gg 1$ a
logarithmic behaviour is attained,
\begin{equation}
C=a+ b \ln  l \;.
\label{eq:C fss log} \end{equation}
 Apart from crossing over to the correct pure result (\ref{eq: C(l)ani})
when $c_{0}\rightarrow 0$, in the Ising model limit,
$(\alpha/\nu)_{\text{pure}} \rightarrow 0$, eq. (\ref{eq:C fss cross})  becomes
 \begin{equation}
C=a+ b \ln (1+ g\ln l )\;,
\label{eq:C fss dotz} \end{equation}
with $g= c_{0} (\alpha/\nu)_{\text{pure}} $.
This is the finite size scaling form which was predicted
analytically\cite{Ising analytic}
and confirmed numerically\cite{Wang1} for the random bond Ising model.

\begin{figure}
\centering
\epsfysize=3.25truein
\epsfxsize=5.truein
\epsffile{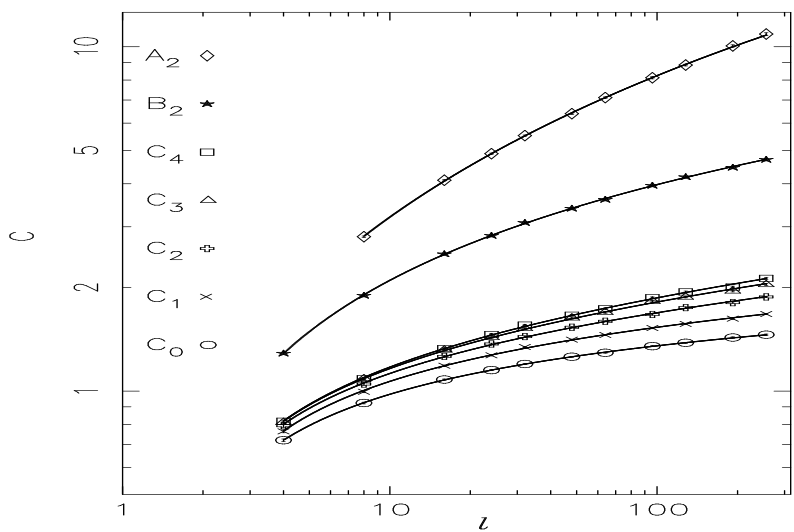}
\caption{ Specific heat, $C$,     as a function of $l$ on a log-log scale
for seven critical RBAT models. $C_{0}$ is a random bond Ising model and
$C_{4}$ is a random bond four state Potts model. The curves are
fits to the form ( \protect\ref{eq:C fss cross}) , yielding estimates for the
coefficients of ( \protect\ref{eq:C fss cross})
 which are listed in Table \ \protect\ref{tab:Cv fit}
 . } \label{fig:CvrCaat} \end{figure}

 In fig. \ref{fig:CvrCaat} the specific heat of the critical RBAT
models is plotted on a log-log scale, with fits to (\ref{eq:C fss cross})
using the full  lattice size range $4\leq l\leq 256$.
The fitting parameters $a_{0},b_{0}$, and $c_{0}$ together with
$(\alpha/\nu)_{\text{pure}}$ and the crossover lengths $l_{c}$ are listed in
Table \ref{tab:Cv fit}. Note that $(\alpha/\nu)_{\text{pure}}$ was not a
fitting
parameter, and was taken for each RBAT model from results of independent
simulations of the corresponding anisotropic AT model.
For the models $C_{0..4}$ (with large randomness, $\frac{ K^{2} }{ K^{1} }
\approx\frac{1}{10}$ ), the crossover lengths $l_{c}$ were found to be 1.
 Nonetheless these models differ by exhibiting some crossover with
 different values of $(\alpha/\nu)_{\text{pure}}$ (see Table  \ref{tab:Cv
fit}). On the other hand, $(\alpha/\nu)_{\text{pure}}$ of the three models
$A_{2},B_{2},C_{2}$ is very similar ( .40, .37, .37 respectively) but they
differ in their amount of randomness,
$\frac{ K^{2} }{ K^{1} } \approx \frac{1}{2},\;\frac{1}{4},\;\frac{1}{10}$
respectively. Consequently, as one would expect, we found that their
crossover lengths decrease as randomness increases:
$l_{c}=51\pm7,\;l_{c}=4.\pm.4,\;l_{c}=1$ respectively.

 \vbox{
 \begin {table}
\caption{
The fitting parameters of the critical specific heat of the random-bond
Ashkin-Teller model. $a_{0},b_{0}$,
and $c_{0}$ were obtained by fitting the specific heat results of the
seven critical RBAT models $C_{0..4}$ and $A_{2}, B_{2}$ to Eq.
( \protect\ref{eq:C fss cross}) using lattice sizes $4\leq l \leq 256$.
$(\frac{\alpha}{\nu})_{\text{pure}}$ is
the specific heat exponent of the corresponding anisotropic (pure) models.
$l_{c}$ is the crossover length defined in ( \protect\ref{eq:C lc}).
Errors are given in parentheses only when error is smaller than or of the
same order as the number itself. }
\label{tab:Cv fit}
\begin{tabular}{llllll}
       & $a_{0}$ & $b_{0}$ & $c_{0}$     &$l_{c}$&$(\frac{\alpha}{\nu})_{
\text{pure}}   $  \\ \hline
$C_{0}$ (Ising)& -.37(12)& .58(1)  & 5.2E4(1.5E4)&  1.    & .0001(150) \\
$C_{1}$        & -4.6    & .51(2)  & 1.5E6       &  1.    & .171(5)    \\
$C_{2}$        & -4.1    & .46(127)& 5.5E4       &  1.    & .375(5)    \\
$C_{3}$        & -3.9    & .43(4)  & 5.5E4       &  1.    & .549(8)    \\
$C_{4}$ (Potts)& -4.1    & .42(1)  & 1.0E5       &  1.    & .630(8)    \\
$B_{2}$        & -.09(5) & 2.00(4) & 1.47(10)    &  4.0(4)& .371(5)    \\
$A_{2}$        & -.07(6) & 9.35(33)&  .26(2)     &  51(7) & .40(1)
 \end{tabular}
\end{table}
}

We found that the magnetization $M$, susceptibility $\chi$ and the
susceptibility of the polarization, $\chi^{(p)}$ are well described at
criticality by the following scaling laws:
\begin{mathletters}  \label{eq:PXp1}
 \begin{equation}
M = A_{M} l^{ -\frac{\beta}{\nu} }   \;,
  \label{eq:M} \end{equation}
 \begin{equation}
\chi = A_{\chi} l^{ \frac{\gamma}{\nu} }   \;,
  \label{eq:Xi} \end{equation}
 \begin{equation}
\chi^{(p)} = A_{\chi^{(p)}} l^{  \frac{\gamma^{(p)}}{\nu} }
  \label{eq:Xp1}\;.  \end{equation}
\end{mathletters}
The estimates for the exponents $\frac{\beta}{\nu}$, $\frac{\gamma}{\nu}$
and
$\frac{\gamma^{(p)}}{\nu}$, which were obtained using lattice sizes $l\geq
24$, are listed in Table \ref{tab: Rand1}.
Even though one observes slight variation of $\frac{\beta}{\nu}$ and
$\frac{\gamma}{\nu}$ from model to model, the results are consistent also
with fixed, non-varying exponents $\frac{\beta}{\nu}=\frac{1}{8}$ and
$\frac{\gamma}{\nu}=\frac{7}{4}$, modified by a logarithmic correction.
So $\frac{\gamma}{\nu}$ shows very little variation or does not vary at all.
This is nearly the same behavior as was found for
the corresponding anisotropic models where $\frac{\gamma}{\nu}$ is predicted
analytically\cite {exact,Nienhuis} to be constant for all models
$\frac{\gamma}{\nu}=\frac{7}{4}$.
The exponent ratio $\frac{\gamma^{(p)}}{\nu}$ connected with the
susceptibility of the polarization which
varies continuously for the pure Ashkin-Teller model seems to do so also for
the random models (see Table \ref{tab: Rand1}).

 \vbox{
 \begin {table}
\caption{  Critical exponents ratios for seven critical RBAT models
$C_{0..4}$ and $A_{2}, B_{2}$. These exponent ratios for the
magnetization $M$, the susceptibility $\chi$ and the susceptibility of the
polarization $\chi^{(p)}$ were obtained by fitting
results for lattice sizes $l\geq 24$ to equation ( \protect\ref{eq:PXp1}).
}
 \label{tab: Rand1}
\begin{tabular}{llll}
       & $\frac{\gamma}{\nu}$ & $\frac{\beta}{\nu}$
& $\frac{\gamma^{(p)}}{\nu}$ \\ \hline
$C_{0}$ (Ising)& 1.751(5)&  .125(3)& 1.549(9)\\
$C_{1}$        & 1.751(6)&  .124(3)& 1.575(8)\\
$C_{2}$        & 1.743(5)&  .129(3)& 1.597(9)\\
$C_{3}$        & 1.736(3)&  .133(2)& 1.638(5)\\
$C_{4}$ (Potts)& 1.714(5)&  .145(3)& 1.714(5)\\
$B_{2}$        & 1.738(4)&  .132(3)& 1.586(6)\\
$A_{2}$        & 1.739(5)&  .132(3)& 1.590(8)
 \end{tabular}
\end{table}
 }

\section{Variances; Definitions and Estimators}  \label{sec: vardef}
In this section we define two types of variances
of thermodynamic quantities in disordered systems and explain their
relation to error estimates.
We explain how the `sample to sample variance' can be estimated from MC
data.

First consider some sample $x$ which is simulated at some temperature $T$.
Because
of the thermal fluctuations and the finite simulation time, we obtain for
this sample an estimate $\overline{P_{x}}$ of the exact $P_{x}$,
with an error
\begin{equation}
 (\delta \overline{P_{x}})^{2} = \frac{\sigma_{T,x}^{2}}
{ T_{MC}/\tau_{x}} \;,     \;\;\;\;\;\;\;\;\;\;T_{MC}\;\;\mbox{  large.}
\label{eq:P Terr}\end{equation}
$T_{MC}$ is the length of the MC runs and $\tau$ is the autocorrelation time
of the MC dynamics. $\sigma_{T,x}^{2}$ is the variance of $P$
within
the sample $x$ due to thermal fluctuations.In practice, in order to avoid the
estimation of $\tau$ which
requires a long simulation time, we estimate $(\delta \overline{P_{x}})^{2}$
by binning the MC sequence into $\sim10$ subsequences and treating each
subsequence as independent ( the Jack-knife procedure).

The estimate for
the error in the estimation of $[\overline{P_{x}}]$, the average of $P$ over
 all samples, is given by
\begin{equation}
 (\delta [\overline{P_{x}}])^{2} =
\frac{1}{(n-1)n}\sum_{x=1}^{n} (\overline{P_{x}}-[\overline{P_{x}}])^2
 \;,     \;\;\;\;\;\;\;\;\;\;n\;\;\mbox{  large,}
\label{eq:P calcerr}\end{equation}
where $n$ is the number of random bond samples. In contrast to (\ref{eq:P
Terr})  this total error has two contributions, namely the sample to
sample fluctuations of the exact $P_{x}$ around $[P_{x}]$ and the thermal
fluctuations of $\overline{P_{x}}$ around $P_{x}$ within each sample,that is
\begin{equation}
(\delta [\overline{P_{x}}])^{2} =
\frac{V}{n}+[ \frac{\sigma^{2}_{T,x}} {n T_{MC}/\tau} ]
 \;, \;\;\;\;\;\;\;\;\;\;n,T_{MC}\;\;\mbox{  large.}
\label{eq:P toterr}\end{equation}
Thus by estimating $(\delta \overline{P_{x}})^{2}$ for all $x$ and $(\delta
[\overline{P_{x}}])^{2}$ with (\ref{eq:P calcerr}) we obtain $V$ through
(\ref{eq:P toterr}); it is an unbiased estimate of the
variance of the exact $P_{x}$ due to sample to sample fluctuations (see
ref. \onlinecite{stat1} for a basic statistical explanation). In order to
minimize the error of $[\overline{P_{x}}]$ for a given amount
of computer time, one needs to adjust $T_{MC}$ so that the two terms in
(\ref{eq:P toterr}) are equal. However, if one is interested in obtaining
a reasonable estimate of $V$, $T_{MC}$ needs to be chosen larger,
so as to obtain accurate estimates of the $\overline{P_{x}}$'s and minimize
the second term on the l.h.s. of (\ref{eq:P toterr}).

As explained in the introduction,the dependence of thermal variance
$\sigma^{2}_{T}$ on the lattice size $l$ has been examined (for
homogeneous models) in ref. \onlinecite{BH:book,self:av}. Thus from here
on the term variance will refer to the variance due to disorder.
Here it is
our aim to examine the dependence of the variance $V$ on $l$ at criticality,
one reason being that for MC simulations of disordered systems,
 it has the bigger influence on their accuracy. This is in addition to the
theoretical motivation given in the introduction. In the next section we
display our results for the variance $V$ of the random-bond Ashkin-Teller
model.
\section{Variance Results of the Random bond Ashkin-Teller models}
\subsection{ Far from Criticality}
Far from criticality the correlation length is finite and one would expect
the system to behave similarly to a collection of independent smaller
systems.
Thus one would expect the Brout argument to hold with the variance scaling
as $l^{-d}$. Nonetheless this is
not obvious: Note that the thermal fluctuations of the specific heat $C$ and
the susceptibility $\chi$ are non self averaging even away from
criticality\cite{BH:book,self:av}.
Thus the RBAT model $C_{2}$ (the choice of model was arbitrary ) was
simulated at the reduced temperature $t=1.$ .
In Fig. \ref{fig: Vat1all} we show
the relative variances $V_{P}/[P]^{2}$, where $V_{P}$ is
defined in (\ref{eq:P toterr}) and $P=E,M,C,\chi$, as a function of
$\log_{10} l$.
 The linear curves are fits to the form $V_{P}/P^{2}=A l^{-\rho}$.
We find $\rho=2.06(7), 2.13(7), 2.04(6), 2.12(7)$ for $\chi,C,E,M$
respectively. Thus the Brout argument is confirmed and far from
criticality strong self averaging holds.

\vbox{
\begin{figure}[h]
\centering
\epsfysize=3.25truein
\epsfxsize=5.truein
\epsffile{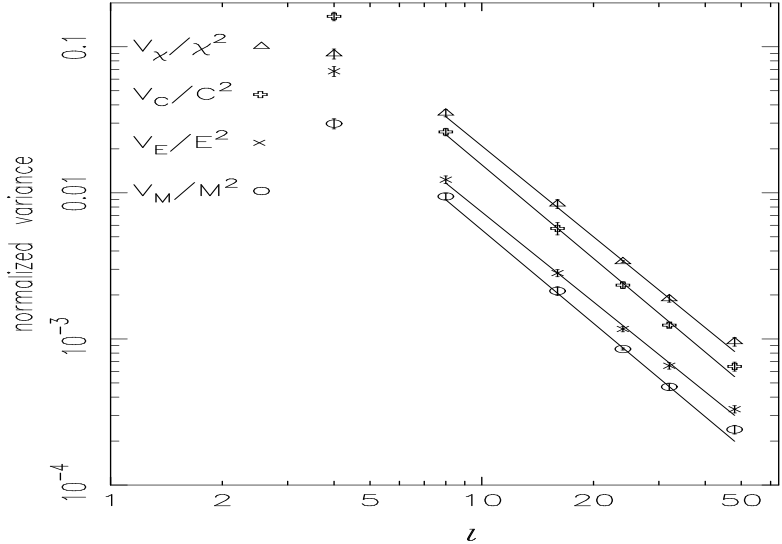}
\caption{ The relative variances $V_{P}/[P]^{2}$, $P=E,M,C,\chi$ as a
function of $\log l$ for the RBAT model $C_{2}$ at the reduced temperature
$t=1.$  The linear curves are fits to the form $V_{P}/P^{2}=A l^{-\rho}$,
yielding $\rho=2.06(7), 2.13(7), 2.04(6), 2.12(7)$ for $\chi,C,E,M$
respectively.
 } \label{fig: Vat1all} \end{figure}
      }

\subsection{ The Variance at Criticality}  \label{sec: res1}
\subsubsection{Distributions}
\begin{figure}
\centering
\epsfysize=3.25truein
\epsfxsize=5.truein
\epsffile{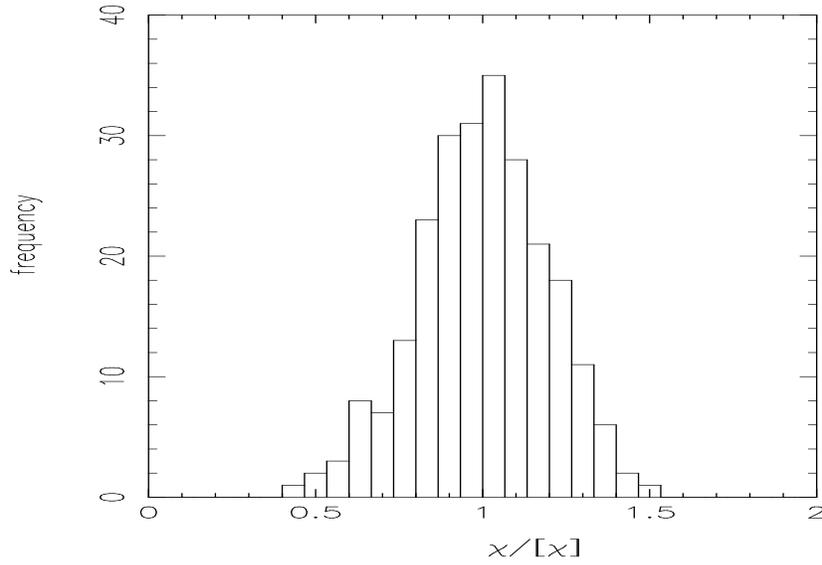}
\caption{Histogram of the frequency of occurrence of samples according to
their susceptibility scaled by the average susceptibility, for the Ising
 model $C_{0}$ and lattice size $l=192$; with 240 samples. }
 \label{fig: HistIsXi} \end{figure}

\begin{figure}
\centering
\epsfysize=3.25truein
\epsfxsize=5.truein
\epsffile{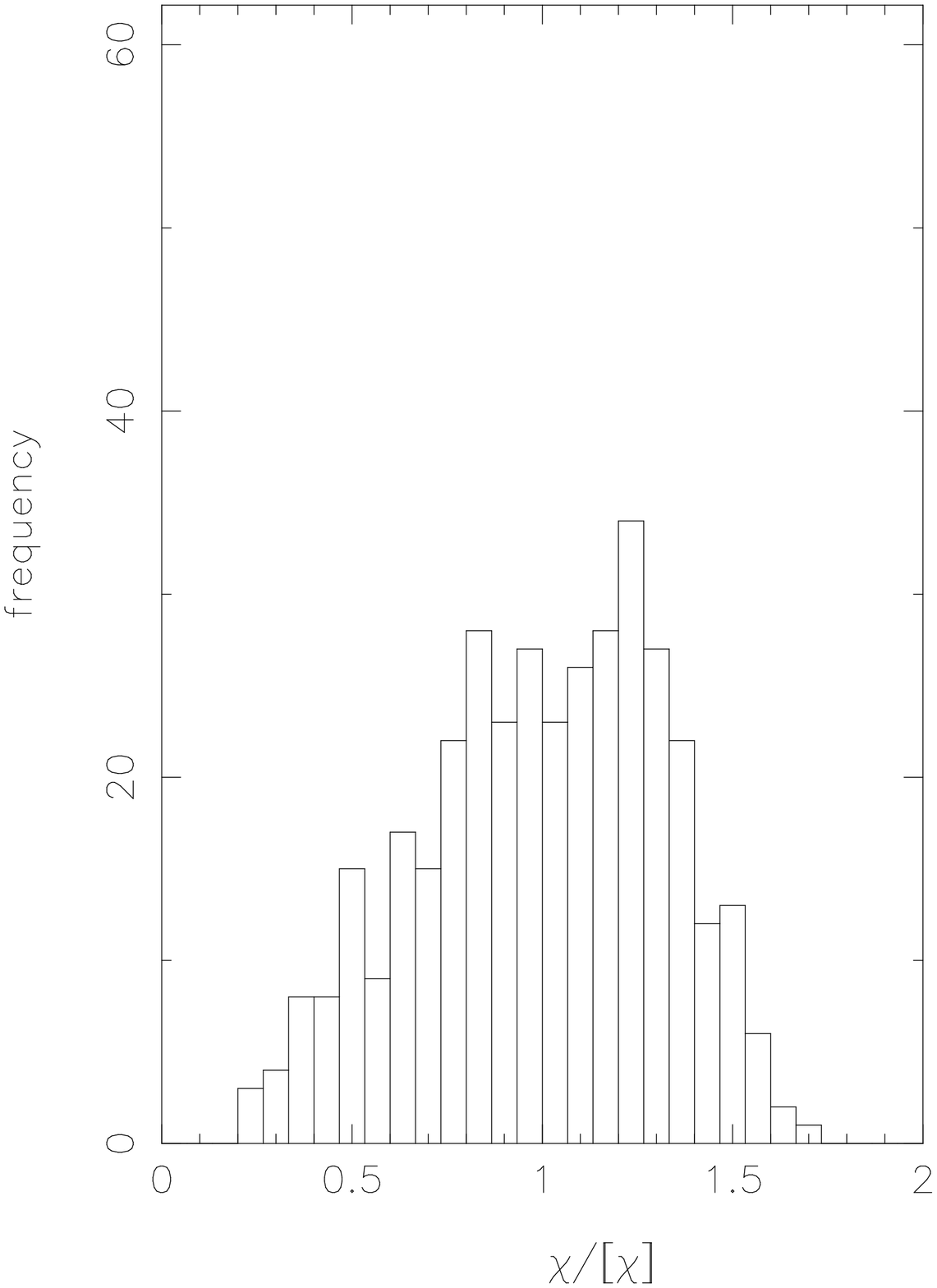}
\caption{Histogram of the frequency of occurrence of samples according to
their susceptibility scaled by the average susceptibility, for the
four-state Potts model $C_{4}$ and lattice size $l=192$; with 370 samples.}
\label{fig: HistPoXi}
\end{figure}

\begin{figure}
\centering
\epsfysize=3.25truein
\epsfxsize=5.truein
\epsffile{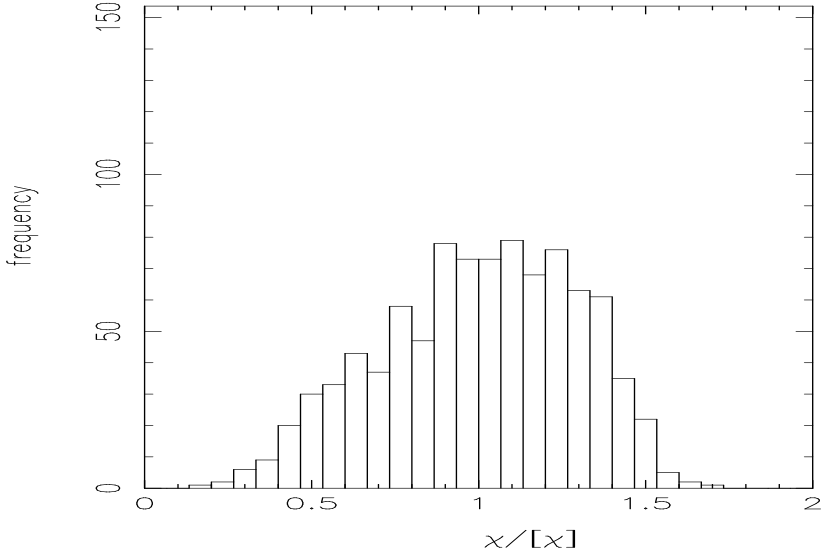}
\caption{Histogram of the frequency of occurrence of samples according to
their susceptibility scaled by the average susceptibility, for the
four-state Potts model $C_{4}$ and lattice size $l=24$; with 920 samples.}
\label{fig: HistPoXi24}
\end{figure}

\begin{figure}
\centering
\epsfysize=3.25truein
\epsfxsize=5.truein
\epsffile{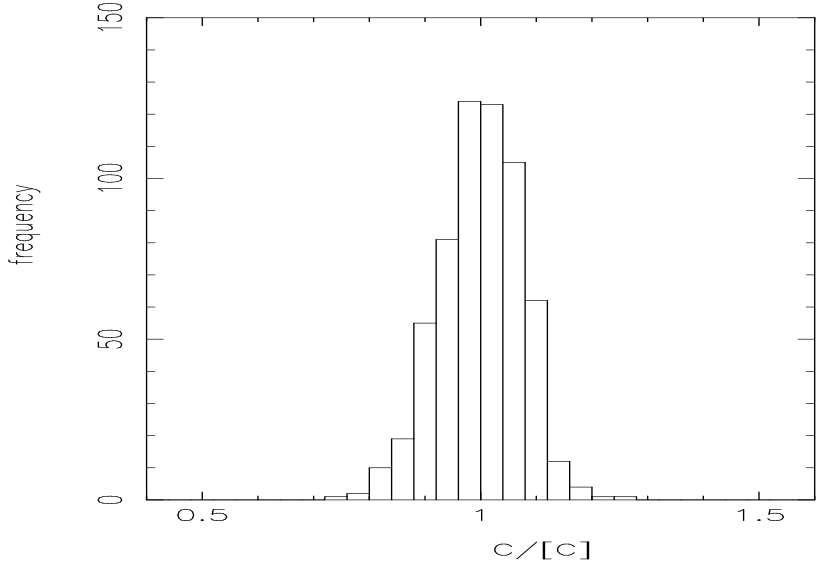}
\caption{Histogram of the frequency of occurrence of samples according to
their specific heat scaled by the average specific heat, for the Ising
 model $C_{0}$ and lattice size $l=48$ with 600 samples. }
\label{fig: HistIsCv}
\end{figure}

\begin{figure}
\centering
\epsfysize=3.25truein
\epsfxsize=5.truein
\epsffile{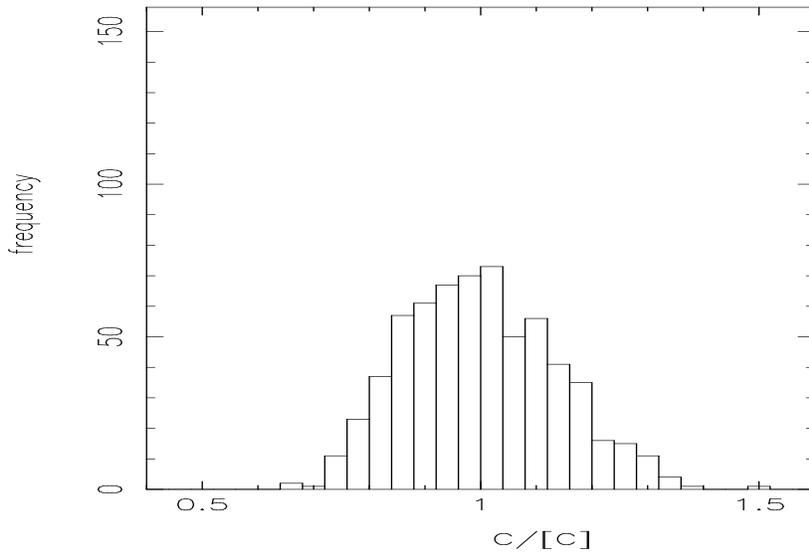}
\caption{Histogram of the frequency of occurrence of samples according to
their specific heat scaled by the average specific heat, for the
four-state Potts model $C_{4}$ and lattice size $l=48$ with 630 samples.}
\label{fig: HistPoCv}
\end{figure}

In order to visualize how large the sample to sample fluctuations are, at
the critical temperature, several histograms of the frequency of occurrence
of samples according
to their susceptibility $\overline{\chi_{x}}$ or according to their specific
heat $\overline{C_{x}}$,
are shown in figures \ref{fig: HistIsXi}-\ref{fig: HistPoCv}. The abscissa
is  scaled by the average susceptibility $[\chi]$ (or specific heat
$[C]$) of all samples.
The histogram of the susceptibility for lattice
size $l=192$ is shown in Fig. \ref{fig: HistIsXi} for the Ising model $C_{0}$
and in Fig. \ref{fig: HistPoXi} for the four-state Potts model $C_{4}$.
The frequency scale of both figures is scaled so that the area of both
histograms is the same. Even though the lattice size is rather large, the
distributions are very wide;
a measurement of a value of $\chi$ at 40\% above the mean
$[\chi]$ has a non-negligible probability for the four state Potts model.

There is a marked difference between the width of the distribution of the
Ising model ($\sqrt{V_{\chi}}/[\chi] \approx 0.2$ ) and the much wider
distribution of the four-state Potts model ($\sqrt{V_{\chi}}/[\chi] \approx
0.32$).
The histogram of the susceptibility for the four-state
Potts model with lattice size $l=24$ is shown in Fig. \ref{fig: HistPoXi24}.
Note that the width of the distribution here is  slightly narrower
($\approx 0.29$) than that of Fig. \ref{fig: HistPoXi}. This very small
difference
 (and even slight increase) of the width as $l$ increases hints at a lack
of self averaging of the susceptibility of the four-state Potts model. An
additional striking difference between the susceptibility distributions of
the four state Potts
model, figures \ref{fig: HistPoXi} and \ref{fig: HistPoXi24}, and the Ising
model, Fig. \ref{fig: HistIsXi}, is that the former are strongly asymmetric
(this asymmetry was measured by measuring the third moment of the
distribution).
A possible explanation for this asymmetry, which exists to some degree in
all the models, is given in Sec. \ref{sec: scale var} and in
\onlinecite{practical} .
The average errors in the estimation of the susceptibility of a single
sample $x$, divided by the average susceptibility,
  $[\delta \overline{\chi_{x}}]/ [\chi]$,
are $0.01, 0.017, 0.012$ in figures \ref{fig: HistIsXi}, \ref{fig: HistPoXi}
and \ref{fig: HistPoXi24} respectively. Since these errors are negligible as
compared to the widths, the histograms are highly reliable.


The histograms of the specific heat for lattice
size $l=48$ are shown in Fig. \ref{fig: HistIsCv} for the Ising model
$C_{0}$
and in Fig. \ref{fig: HistPoCv} for the four-state Potts model $C_{4}$.
Note that the distributions of the specific heat are  much narrower
than those of the susceptibility.
The width of the distribution for the four-state Potts model ($\approx 0.126$)
is about twice wider than the width of
the distribution for the Ising model ($\approx .062$).
The asymmetry of the distribution for the four-state Potts model
is almost unnoticeable and is of the opposite sign than the asymmetry
 of the susceptibility.
The average error in the estimation of the specific heat of a single sample
$x$ divided by the average specific heat $[\delta \overline{C_{x}}]/[C]$ is
$0.038 $ for the Ising model and $0.048$ for the four-state Potts model, so
that these histograms are much less accurate than those of the
susceptibility. For the larger lattices the ratio between the width and the
error becomes smaller, mostly because the width becomes smaller, and
histograms become even less accurate. Thus in order to obtain more accurate
histograms and also better estimates of the variance (which is the
square of the width of the histograms), longer simulation times would be
needed, in order to obtain more accurate estimates of the
$\overline{C_{x}}$'s. This may be done in a future study.


\subsubsection{The variance}

\begin{figure}
\centering
\epsfysize=3.25truein
\epsfxsize=5.truein
\epsffile{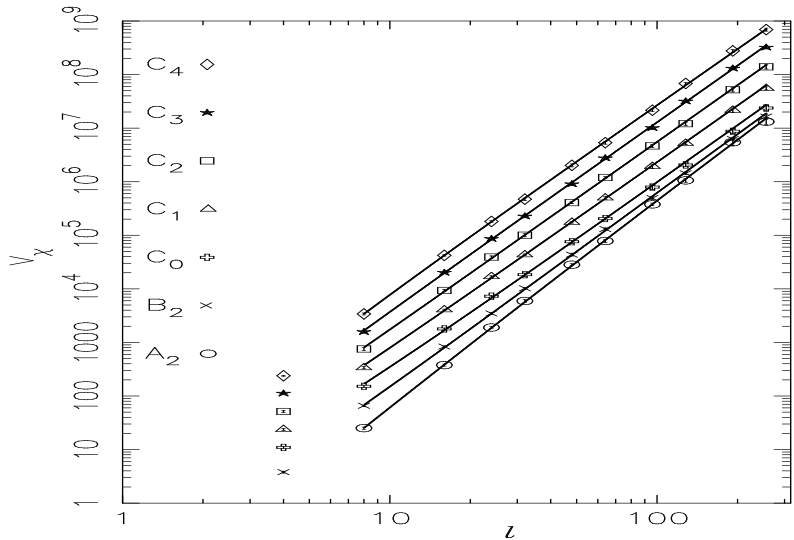}
\caption{ The variance of $\chi$, $V_{\chi}$ as a function of $\log l$ for
all critical
models, $C_{0..4}$, and $A_{2},B_{2}$ of the RBAT model.
For the sake of
clarity $V_{\chi}$ of the model $C_{i}$ was multiplied by $2^{i+1}$.
The solid lines are  fits to the
form ( \protect\ref{eq: var RBAT}) , yielding estimates for fitting
parameters which are listed in table \ \protect\ref{tab: var}
. } \label{fig: XtVfit2} \end{figure}

In Fig. \ref{fig: XtVfit2} we show the variance of $\chi$, $V_{\chi}$ of the
seven critical RBAT models. For the sake of clarity (so that the data do not
fall on top of each-other) $V_{\chi}$ of the model $C_{i}$ was multiplied by
$2^{i+1}$.
The lines are fits according to a theory which we develop in the next
section. Here we just note that $V_{\chi}$ is measured with high precision,
so that it may be faithfully tested against theory.

\begin{figure}
\centering
\epsfysize=3.25truein
\epsfxsize=5.truein
\epsffile{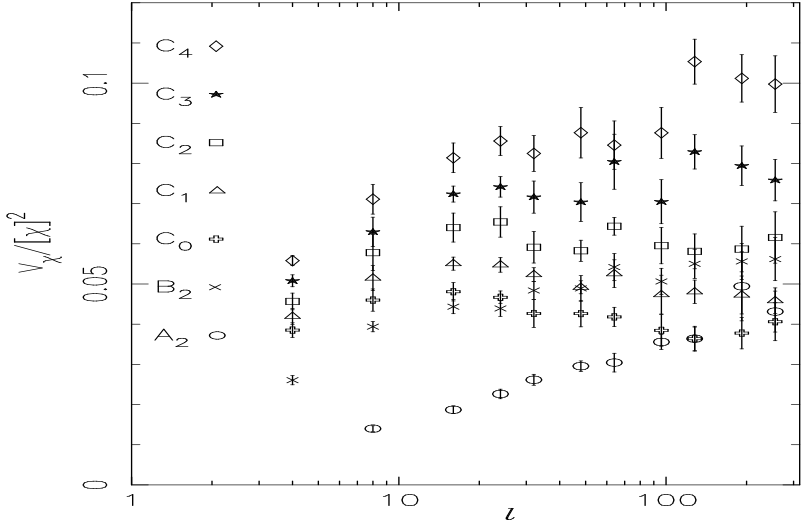}
\caption{ The scaled variance of the susceptibility, $V_{\chi}/[\chi]^{2}$
as a function of $\log l$ for all critical
models, $C_{0..4}$, and $A_{2},B_{2}$ of the RBAT model.
 } \label{fig: XtVNor2} \end{figure}

The relative variance
$V_{\chi}/[\chi]^{2}$ is plotted in Fig. \ref{fig: XtVNor2}. Since it is the
ratio of two fluctuating quantities, the errors are quite large. Nonetheless
the main trends can be seen.
First note that apparently for all models (except for
the weakly random model $A_{2}$) $V_{\chi}/[\chi]^{2}
\rightarrow \text{const}$, so that the susceptibility is non-self averaging.
 It is also possible that $V_{\chi}/[\chi]^{2}$ is slightly increasing with
$l$ for some models ( e.g. the four state Potts model $C_{4}$ ) or slightly
decreasing for the Ising model $C_{0}$.
Upon comparison of the models $C_{i} \; i=0\ldots 4$ we make the
following observations. The higher is the specific heat of a model, the
larger is its relative variance (see Fig. \ref{fig:CvrCaat}). The higher is
the exponent
$(\frac{\alpha}{\nu})_{\text{pure}}$ of the pure model (see Table
\ref{tab:Cv fit}), the larger is
the initial slope of the relative variance of the corresponding random model.
Thus the relative variance of the Ising model $C_{0}$ is the smallest and
the
increase with $l$, for small $l$, is the smallest. The relative variance
of the four-state Potts model $C_{4}$ is the largest and the increase with
$l$, for small $l$, is the largest. The relative variances of the RBAT
models $C_{1,2,3}$ fall in between.
The relative variance of the weakly random model $A_{2}$ shows a steady
 increase with $l$, in contrast with the highly random model $C_{2}$, in
which a shorter increase is followed by a plateau. This is reminiscent of the
specific heat of the $A_{2}$ model which exhibits very slow crossover from
the power-law behavior (\ref{eq: C(l)ani}) to the
asymptotic logarithmic behavior (\ref{eq:C fss log}) with a crossover length
 of $l_{c}\approx50$.
Thus for small lattice sizes the model $A_{2}$ exhibits effective
exponents (of the specific heat) of the pure model, and also exhibits a
small variance due to
its small degree of randomness. But as the lattice size increases, this
effect
diminishes, the effective exponents approach the random value, and the
variance approaches that of the highly random models.

\begin{figure}
\centering
\epsfysize=3.25truein
\epsfxsize=5.truein
\epsffile{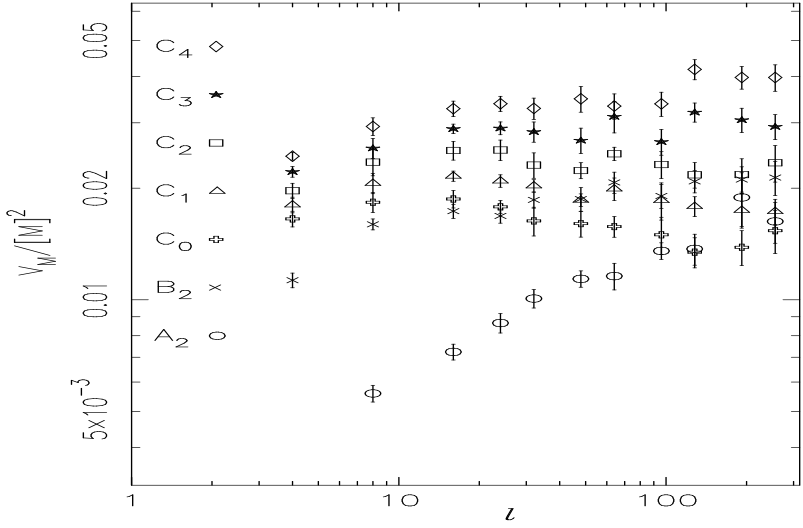}
\caption{ The scaled variance of $M$, $V_{M}/[M]^{2}$ as a
function of $\log l$ for all critical
models, $C_{0..4}$, and $A_{2},B_{2}$ of the RBAT model.
 } \label{fig: M1VNor2} \end{figure}

A very similar picture is obtained for the
relative variance of the magnetization $V_{M}/[M]^{2}$, as seen in figure
\ref{fig: M1VNor2}. The qualitative picture of the
magnetization results, Fig. \ref{fig: M1VNor2}, is very similar to that of
the susceptibility results, Fig. \ref{fig: XtVNor2}, showing the same trends
as outlined above. Yet we emphasize that even though the magnetization is
an intensive quantity, it does not seem that $V_{M}/[M]^{2}\rightarrow 0$
as $l$ increases so that the magnetization is {\em not } self averaging at
criticality.

\begin{figure}
\centering
\epsfysize=3.25truein
\epsfxsize=5.truein
\epsffile{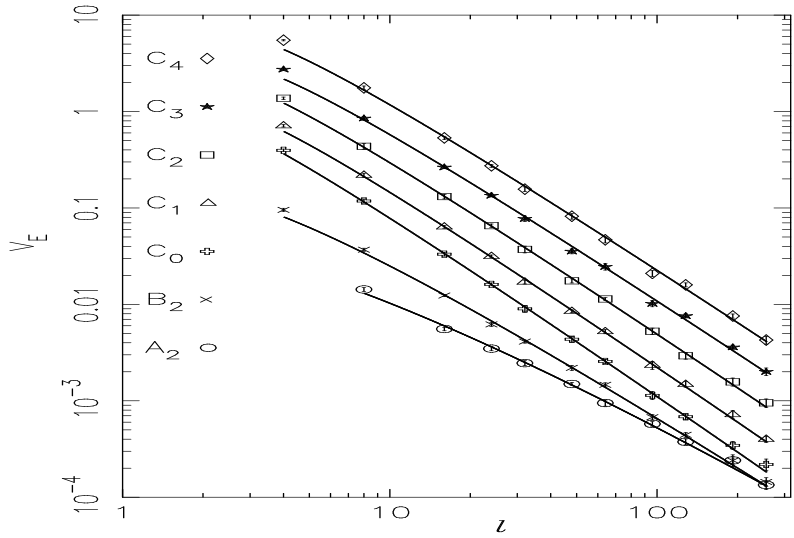}
\caption{ The variance of the energy, $V_{E}$ as a function of $\log l$ for
all critical models, $C_{0..4}$, and $A_{2},B_{2}$ of the RBAT model.
For the sake of
clarity $V_{E}$ of the model $C_{i}$ was multiplied by $2^{i+1}$.
The solid lines are fits to
the form ( \protect\ref{eq: varE RBAT}) , yielding estimates for the fitting
parameters $a_{v},b_{v}$ which are listed in table \ \protect\ref{tab: var}
. } \label{fig: EnRVft2} \end{figure}

In Fig. \ref{fig: EnRVft2} the variance of the energy
$V_{E}$\cite{E not nor} is plotted
on a log-log scale. For the sake of clarity (so that the data do not fall on
top of each-other) $V_{E}$ of the model $C_{i}$ was multiplied by $2^{i+1}$ .
We fit the data to the
form $V_{E}\sim l^{-\theta}$ for lattice sizes $l\geq16$ (but the fitting
curves shown in Fig. \ref{fig: EnRVft2} are not made with this form but
with a more complicated one which is due to a theory which we develop in
the next section). The highest
value of $\theta$, $\theta=1.855(13)$, was obtained for the Ising model
$C_{0}$. For the four-state Potts model, $C_{4}$, we obtained
$\theta=1.72(2)$, while for the models $C_{1,2,3}$ the values of $\theta$
fell between these two values. Thus in contrast with the susceptibility and
the magnetization, the variance of the energy $V_{E}$ is weakly
self-averaging.
But similar to the susceptibility, models with a higher specific heat or
with a higher effective $\frac{\alpha}{\nu}$ have a smaller $\theta$, and
thus their $V_{E}$ decreases more slowly with $l$.
For the weakly random model $A_{2}$, $\theta=1.30(3)$ so that again the
slope of the variance is correlated with the high slope of the specific heat
of this model.

\begin{figure}
\centering
\epsfysize=3.25truein
\epsfxsize=5.truein
\epsffile{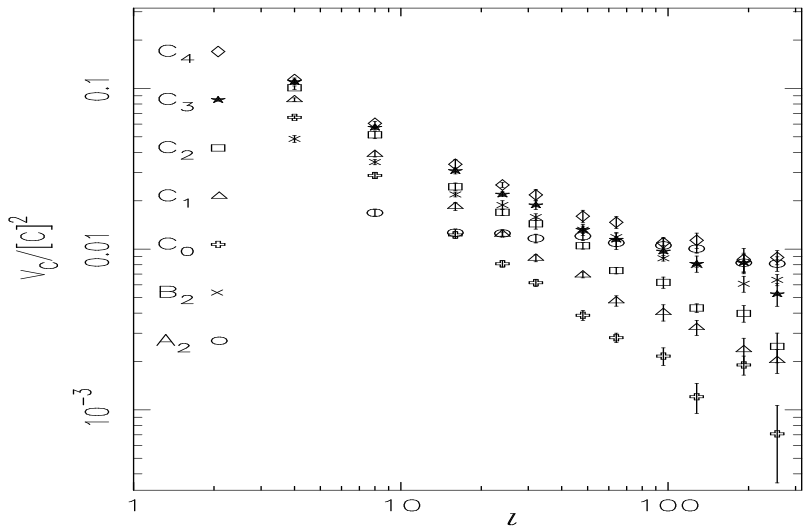}
\caption{ The scaled variance of $C$, $V_{C}/[C]^{2}$ as a
function of $\log l$ for all critical
models, $C_{0..4}$, and $A_{2},B_{2}$ of the RBAT model.
 } \label{fig: CvVNor2} \end{figure}

The  results of the relative variance $V_{C}/C^{2}$, plotted in Fig.
\ref{fig: CvVNor2}, seem to indicate that the specific heat is weakly self
averaging. Nonetheless the effective slopes increase with $l$ (or the
absolute values of the slopes decrease with $l$,
 this trend being strongest for the four-state Potts model $C_{4}$ ) so that
it is possible that self averaging does not hold for very large $l$. It also
seems possible that the Ising model is self-averaging while the other models
are not. Clearly more accurate data and data from larger systems would be
useful.
As in other variances, we observe qualitatively that the relative variance
of the
moderately random models, $A_{2}$ and $B_{2}$, approaches that of
the highly random ones as $l$ increases and even exceeds it. \\
The findings of this section are partly summarized in table \ref{tab:
summary}, where the self averaging properties of the highly random critical
 models are displayed.

\vbox{
 \begin {table}
\caption{  Summary of the self averaging properties of the critical
random-bond four-state Potts $C_{4}$, Ashkin-Teller $C_{1..3}$, and
Ising $C_{0}$ models. The letter `n' stands for non self averaging, `w' for
weakly self averaging, and `?' for inconclusive results. This summary is
according to a subjective examination of the numerical results
as displayed in figures
\protect\ref{fig: XtVNor2} , \protect\ref{fig: M1VNor2} , \protect\ref{fig:
EnRVft2} , \protect\ref{fig: CvVNor2} \protect\ref{fig: XPVfit2} . According
to our theory only the
energy $E$ is weakly self averaging while all other quantities are non self
averaging in all of these models. If the theory is correct then other
behavior implied by the numerical data are merely transients. }
 \label{tab: summary}
\begin{tabular}{llllll}
model                     & $\chi$ & $\chi^{(p)}$ & $M$ & $C$  & $E$ \\
\hline
$C_{0}$ (Ising)           & ?      &  ?           & ?   &  w   &  w  \\
$C_{1..3}$ (Ashkin-Teller)& n      &  n           & n   &  w   &  w  \\
$C_{4}$ (4 state Potts)   & n      &  n           & n   &  w   &  w  \\
 \end{tabular}
\end{table}
}

In the next section we develop a phenomenological finite size scaling theory
for the variance. This theory explains
the apparent connection between the variance and the specific heat
behavior of the random models. In the last section we explain how this
theory was applied to the results we have displayed here and discuss the
comparison between our scaling theory and the numerical results.

\section{ Finite Size Scaling of Sample to Sample Fluctuations at
Criticality } \label {sec: scale var}
As our numerical results show, we have obtained quite accurate estimates of
the
variance $V$ of the thermodynamic functions at the critical temperature for
different lattice sizes $l$.
In order to understand these results, a phenomenological theory of finite
size scaling of disordered systems, which will take into account sample to
sample fluctuations, needs to be developed.

The main result of our theory will be the scaling of the variance $V$ with
$l$ at criticality.
 To be precise, we will calculate the variance $V$ of $P$ (e.g. $P=C,M,E$ or
 $\chi$, where all these quantities are normalized per volume; i.e. they are
densities)
\begin{equation}
 V(T,l)=[ ( P_{x}(T,l) - [ P_{x}(T,l) ])^{2} ] \;.
\label{eq: V(T,l)} \end{equation}
 $P_{x}(T,l)$ is the exact value of $P$ (that is, after the thermal
fluctuations have been averaged over) of a specific sample $x$ (with
some specific realization of randomly distributed bonds) of linear size $l$
at temperature $T$. Again
the square brackets denote averaging over the different samples $x$.

Our conclusion will be that when the specific heat exponent $\alpha$ is
negative the leading behavior of $V$ at $T_{c}$ is
\begin{equation}
  V(T_{c},l)\sim K^{2}_{r}l^{2\rho+\frac{\alpha}{\nu} } \;,\;\;\;\;
 \text{ implying } \;\;\;\;\;
 \frac{ V(T_{c},l) }{ [ P_x (T_{c},l) ]^2 } \sim l^{ \frac{\alpha}{\nu} }
 \;. \label{eq: Vlead} \end{equation}
Where $K_{r}$ is a measure of the amount of randomness or disorder and
$\rho$ is the critical exponent of the quantity $P$, e.g. if
$P=\chi$ then $\rho=\frac{\gamma}{\nu}$.
Eq. (\ref{eq: Vlead}) implies that disordered systems
at criticality are only weakly self averaging when $\frac{\alpha}{\nu}<0$.
For $\frac{\alpha}{\nu}=0_{+}  $ (log), as was found\cite{RBAT} for the
random
bond Ashkin-Teller model,  our derivation is strictly not valid for $l\gg 1$.
Nonetheless for the range of lattice sizes considered, we found
good agreement between the numerical results for the variance of
$\chi,\chi_{p}$ and $E$ and theoretical fits according to (\ref{eq: Vlead})
together with next to leading terms (see figures \ref{fig: XtVfit2} ,
\ref{fig: XPVfit2} and \ref{fig: EnRVft2} and discussion in the next
section). If no dramatic change occurs at larger sizes, then the sample
to sample fluctuations of the random bond Ashkin-Teller model
are non self averaging.

The result (\ref{eq: Vlead}) indicates that the sample to sample
fluctuations at the
critical temperature $T_{c}$ depend strongly on the specific heat exponent
$\frac{\alpha}{\nu}$. This strong dependence can be made plausible based on
heuristic arguments. These heuristic arguments will serve to define some
basic ingredients of our approach and will be followed by a more
quantitative treatment.

 We start by characterizing every
  specific sample $x$ of size $l$ by a pseudo-critical
temperature $T_{c}(x,l)$. This pseudo-critical temperature can, for
instance, be the temperature at which a maximum in the specific heat of
the sample occurs. We denote the average pseudo-critical temperature as
$T_{c}(l)=[ T_{c}(x,l) ]$. We assume that, as is the case in homogeneous
systems,
 \begin{equation}
 T_{c}(l)-T_{c}=a l^{-y_{t}}
\;, \label{eq: Tc scaling} \end{equation}
 where $a$ is a constant, $y_{t}=1/\nu$ and $T_{c}=
\lim_{l\rightarrow\infty}
T_{c}(l)$. $T_{c}$ is the average critical temperature of the ensemble of
infinite samples.  Eq. (\ref{eq: Tc scaling}) is supported by a numerical
study\cite{Hen} of the three dimensional dilute Ising model.

Next we assume that $T_{c}(x,l)$ fluctuates around $T_{c}(l)$ with width
 \begin{equation}
 \delta T_{c}(l)\sim l^{-d/2}
 \;. \label{eq: width} \end{equation}
This assumption is probably true\cite{Stinch,Harris} for small
disorder and small $l$, or close to the pure system fixed point.
 We assume it without proof, for large disorder as well, though
 for large disorder (or close to the random fixed point) the possibility
that $\delta T_{c}(l)\sim l^{-y_{t}} $  has been raised\cite{private}. \\
Define reduced temperatures
 \begin{mathletters} \label{eq:def reduce}
 \begin{equation}
t_{c}(x,l)=\frac{ T_{c}(x,l)-T_{c} }{T_{c}}
\;,\label{eq:def reduce a} \end{equation}
 \begin{equation}
t_{c}(l)=\frac{ T_{c}(l)-T_{c} }{T_{c}}
\;,\label{eq:def reduce b} \end{equation}
and the reduced width
 \begin{equation}
\delta t_{c}(l)=\frac{ \delta T_{c}(l) }{T_{c}}
\;.\label{eq:def reduce c} \end{equation}
\end{mathletters}
We make (\ref{eq: width}) more specific by
assuming for $t_{c}(x,l)$ a Gaussian probability distribution $q(t_{c}(x,l))$
\begin{equation}
  q(t_{c}(x,l))=\frac{ l^{d/2} }{ \sqrt{2\pi} K_{r} } \exp\{
-[t_{c}(x,l)-t_{c}(l)]^{2} l^{d} /2K^{2}_{r} \}
 \;. \label{eq: prob} \end{equation}
The width of the distribution is controlled by the lattice size $l$ and by
$K_{r}$ which is a measure of the amount of randomness or disorder.

The scaling relations (\ref{eq: Tc scaling}, \ref{eq:
width}) already make the result (\ref{eq: Vlead}) plausible.
The main idea is that the sample to sample fluctuations at
$T_{c}$ are governed by the relative magnitude of two temperature
differences. The first is the difference between the average
pseudo-critical temperature $T_{c}(l)$ and the critical temperature of the
infinite system $T_{c}$. The second is the difference between $T_{c}(l)$
and $T_{c}(x,l)$, the pseudo-critical temperature of the sample $x$, which
is governed by $\delta t_{c}(l)$. If $\delta t_{c}(l)\gg |t_{c}(l)|$
then fluctuations in $t_{c}(x,l)$ are so large that for some samples one will
find $T_{c}>T_{c}(x,l)$ while for other samples $T_{c}<T_{c}(x,l)$. In this
case, even though we are simulating all samples at $T_{c}$, some
samples are in their high temperature phase while others
are in their low temperature phase. This will obviously increase the
sample to sample fluctuations in any observable. If, on the other hand,
 $\delta t_{c}(l)\ll |t_{c}(l)|$,  then $T_{c}-T_{c}(x,l)$ will always have
the same sign
and fluctuations will be smaller. The condition $\delta
t_{c}(l)\ll |t_{c}(l)|$
will be fulfilled for large $l$ if $y_{t}-\frac{d}{2}<0$ or, using the
hyper-scaling relation $\frac{\alpha}{\nu}=2y_{t}-d$, if
$\frac{\alpha}{\nu}<0$. For disordered systems, the bound $y_{t}\leq d/2$
has been proven by Chayes et al.\cite{Chayes}, so that asymptotically one
always finds
$\frac{\alpha}{\nu}\leq 0$. However, for small $l$ and small disorder, the
system may be governed by a positive $(\frac{\alpha}{\nu})_{\text{pure}}$.
In this case sample to sample fluctuations can increase with lattice size,
as is indeed seen in our numerical results for the weakly disordered model
$A_{2}$.
Thus on the basis of these considerations one can
conclude that the sign and magnitude of the specific heat exponent
$\alpha$ of the disordered model have a strong influence on the sample to
sample fluctuations \cite{practical}, and will determine whether they are
self averaging. The discussion above is analogous to the physical
arguments leading to the Harris criterion\cite{Harris}, but in a finite size
scaling formulation. The difference is that the Harris criterion was
derived near the pure system fixed point, while we are assuming
that similar conditions apply also next to the disordered critical fixed
point.

In order to put these general considerations on more quantitative grounds,
we proceed to derive the finite-size scaling expression (\ref{eq: Vlead})
for the variance of various thermodynamic quantities. Start by introducing
the reduced temperature of each sample $x$;
 \begin{equation}
\dot{t}_{x}=\frac{T-T_{c}(x,l)}{T_{c}}
\;. \label{eq: tx dot} \end{equation}
We assume ( for samples with $T$ close to $T_{c}(x,l)$) a finite size
scaling form for the singular part of $P_{x}$,
 \begin{equation}
  P^{\text{sing}}_{x}(T,l)= l^{\rho} \tilde{Q}_{x}( \dot{t}_{x}l^{y_{t}} )
\;. \label{eq: Psing} \end{equation}
The form of the function $\tilde{Q}_{x}(Z)$ (or its coefficients) are
assumed to be sample dependent but the critical exponents $\rho,y_{t}$ are
assumed to be universal or sample independent.

eq. (\ref{eq: Psing}) embodies the usual\cite{usual} finite size scaling
assumption
that in the vicinity of the critical temperature the behaviour of a large
finite system is governed by the scaled variable $\xi_{x}/l$. We use this
assumption, even though in the present context it implies that a single
correlation length $\xi_{x}$ is sufficient
to describe the state of a disordered sample, which is not obvious at all.
This ``thermal'' $l$ -dependence is compounded by the fact that if we
increase
$l$, we must generate additional random bonds, and hence increasing $l$
necessitates, effectively, changing $x$ (that represents a particular
realization of the random bond variables). Since $x$ affects $P^{sing}_{x}$
through the non-universal coefficients of $\tilde{Q}_{x}$, a non-thermal
dependence of $P^{sing}_{x}$ on $l$ is induced. The main task of our
analysis is to separate the thermal $l$ dependence from the non-thermal
component.

 At this stage it is possible to draw some more conclusions based on
 (\ref{eq: Psing}), without making strong assumptions about the coefficients
of $\tilde{Q}_{x}$. We leave such derivations for the Appendix. Here we
proceed in a more straightforward manner by using a simplifying ansatz.
Our ansatz states that the coefficients of
$\tilde{Q}_{x}$ depend only on $\Delta t_{c}(x,l)$, the deviation of the
pseudo-critical temperature of the sample from the
 average pseudo-critical temperature, defined as
 \begin{equation}
\Delta t_{c}(x,l)= t_{c}(x,l) - t_{c}(l)
\;. \label{eq: Delta} \end{equation}

It is convenient to proceed by rewriting $\dot{t}_{x}$ as
$\dot{t}_{x}=t-\Delta t_{c}(x,l)-t_{c}(l)$ with
$t=\frac{T-T_{c}}{T_{c}}$. Using the scaling of $t_{c}(l)$ [see (\ref{eq: Tc
scaling}) and (\ref{eq:def reduce b})] , we substitute $\tilde{Q}_{x}$ by a
different scaling function $Q_{x}$ and rewrite Eq. (\ref{eq: Psing}) as
\begin{equation}
P^{sing}_{x}(T,l)= l^{\rho} \tilde{Q}_{x}\{ (t-\Delta t_{c}(x,l)-t_{c}(l))
 l^{y_{t}} \}
		   = l^{\rho} Q_{x}\{ (t-\Delta t_{c}(x,l))l^{y_{t}} \}
\;. \label{eq: Psing2} \end{equation}
  For completeness of the treatment which will later prove to be necessary
we do not neglect the analytic dependence of $P_{x}(T,l)$ on
$(t-\Delta t_{c}(x,l))$\cite{analytic}, and write
  \begin{equation}
  P_{x}(T,l)= A_{x}+B_{x}(t-\Delta t_{c}(x,l))+C_{x}(t-\Delta t_{c}(x,l))^{2}
  +\ldots+l^{\rho} Q_{x}\{ (t-\Delta t_{c}(x,l))l^{y_{t}} \}
\;. \label{eq: scaling1} \end{equation}
 The coefficients $A_{x},B_{x},C_{x}$ are
assumed to be sample dependent in the same way that the coefficients of
$Q_{x}$ are; namely they depend only on $\Delta t_{c}(x,l)$\cite{Tc(x)}.
Next, assume the dependence of the coefficients on $\Delta t_{c}(x,l)$ is
analytic. Since according to (\ref{eq: prob}) and (\ref{eq: Delta}),
$\Delta t_{c}(x,l)$ is distributed around
zero with width that scales as $l^{-d/2}$, we can expand
 \begin{equation}
  A_{x}=A_{0}+A_{1}\Delta t_{c}(x,l)+A_{2}(\Delta t_{c}(x,l))^{2}+\ldots
 \;, \label{eq: Ax} \end{equation}
 where $A_{0},A_{1},A_{2}$ are sample independent.
 The same type of expansion is assumed for $B_{x},C_{x}$ etc.

We are interested in knowing what happens at $T=T_{c}$, the average critical
temperature of the ensemble of infinite samples. Thus we set
$T=T_{c}$ which implies $t=0$. For the analytic part of
(\ref{eq: scaling1}) we get
\begin{eqnarray}
  P_{x}^{\text{analytic}}(T_{c},l)= & A_{x}-B_{x}\Delta t_{c}(x,l)+
		      C_{x}\Delta t_{c}(x,l)^{2}+\ldots=   \nonumber \\  &
A_{0}+(A_{1}-B_{0})\Delta t_{c}(x,l)+(A_{2}-B_{1}+C_{0})
	    (\Delta t_{c}(x,l))^{2}+\ldots   \nonumber \\  &
\equiv a + b \Delta t_{c}(x,l) + c (\Delta t_{c}(x,l))^{2}
\;, \label{eq: anal Tc} \end{eqnarray}
where the second equality is reached by use of (\ref{eq: Ax}) and the
same expansions for other coefficients. The last equality is a
redefinition of constants.
In a similar way we expand $Q_{x}$
 \begin{equation}
 Q_{x}(Z)= D_{x}+E_{x}Z+F_{x}Z^{2}+\ldots
 \;, \label{eq: Q(X)} \end{equation}
 where $D_{x},E_{x},F_{x}$ are again expanded as in (\ref{eq: Ax}).
 Again setting $t=0$, we obtain for the singular part of
(\ref{eq: scaling1})
 \begin{equation}
  P_{x}^{singular}(T_{c},l)=l^{\rho}\{ D_{0}+( D_{1}-E_{0}l^{y_{t}}
)\Delta t_{c}(x,l) + ( D_{2}-E_{1}l^{y_{t}}+F_{0}l^{2y_{t}} )
(\Delta t_{c}(x,l))^{2} \}+\ldots
\;. \label{eq: sing Tc} \end{equation}
We stress that since we set $Z=-\Delta t_{c}(x,l) l^{y_{t}}$, $Z$ is
fluctuating around zero with width that scales as
$\sim l^{\frac{\alpha}{2\nu}}$. Thus
the expansion (\ref{eq: Q(X)}) is justified asymptotically only for
$\alpha<0$. Putting together (\ref{eq: anal Tc}) and (\ref{eq: sing Tc})
we have
 \begin{eqnarray}
  P_{x}(T_{c},l)= & ( a+D_{0}l^{\rho} )+
( b+D_{1}l^{\rho}-E_{0}l^{\rho+y_{t}} )\Delta t_{c}(x,l) + \nonumber \\ &
( c+ D_{2}l^{\rho}-E_{1}l^{\rho+y_{t}}+F_{0}l^{\rho+2y_{t}} )
(\Delta t_{c}(x,l))^{2} +\ldots
\equiv d +e \Delta t_{c}(x,l)+f (\Delta t_{c}(x,l))^{2}
\;. \label{eq: Px Tc} \end{eqnarray}
Notice that here the only dependence on the specific sample $x$ is through
explicit dependence on $\Delta t_{c}(x,l)$, the deviation of its reduced
pseudo-critical temperature from the average pseudo-critical temperature.
Taking the quenched sample average [ ] with the probability distribution
(\ref{eq: prob}), using $[\Delta t_{c}(x,l)]=0$, we get
 \begin{equation}
  [P_{x}(T_{c},l)]= d+f [(\Delta t_{c}(x,l))^{2}]
 \;, \label{eq: Pav} \end{equation}
 and using $[(\Delta t_{c}(x,l))^{3}]=0$ we further obtain
 \begin{equation}
 [(P_{x}(T_{c},l))^{2}]= d^{2}+ e^{2}[(\Delta t_{c}(x,l))^{2}]+ f^{2}
[(\Delta t_{c}(x,l))^{4}]  +2df[(\Delta t_{c}(x,l))^{2}]
 \;. \label{eq: P2av} \end{equation}
The variance is then given by
 \begin{equation}
  V(T_{c},l)=
    e^{2}[(\Delta t_{c}(x,l))^{2}]+ f^{2} \{ [(\Delta t_{c}(x,l))^{4}] -
[(\Delta t_{c}(x,l))^{2}]^{2} \}
    =e^{2}[(\Delta t_{c}(x,l))^{2}]+ 2f^{2} [(\Delta t_{c}(x,l))^{2}]^{2}
 \;, \label{eq: Vav} \end{equation}
 where the last equality is a property of the Gaussian distribution.
 Lastly we use $[(\Delta t_{c}(x,l))^{2}]=K^{2}_{r}l^{-d}$ and obtain to the
leading orders in $l$
 \begin{eqnarray}
  V(T_{c},l)=   &
( b^{2}+D_{1}^{2}l^{2\rho}+E_{0}^{2}l^{2\rho+2y_{t}}+
2bD_{1}l^{\rho}-2bE_{0}l^{\rho+y_{t}}-2D_{1}E_{0}l^{2\rho+y_{t}} )
K^{2}_{r}l^{-d}+ \nonumber \\ &
( F_{0}^{2}l^{2\rho+4y_{t}} )2K^{4}_{r}l^{-2d} +\ldots
 \;. \label{eq: Vav(l)} \end{eqnarray}
Since $y_{t}>0$, and usually $\rho+y_{t}>0$, the leading term in (\ref{eq:
Vav(l)}) is
 \begin{equation}
  V(T_{c},l)\sim
E_{0}^{2}K^{2}_{r}l^{2\rho+2y_{t}-d}=
E_{0}^{2}K^{2}_{r}l^{2\rho+\frac{\alpha}{\nu} }
 \;. \label{eq: Vlead2} \end{equation}
The last term in (\ref{eq: Vav(l)}) is proportional to
$K^{4}_{r}l^{2\rho+2\frac{\alpha}{\nu}}$, and may be neglected with
respect to (\ref{eq: Vlead2}) only for $\frac{\alpha}{\nu}<0$, or if
$K^{2}_{r}\ll1$ and $l$ is not too large.  (\ref{eq:
Vlead2}) is our main result for the variance, where all exponents
$\rho,\alpha,\nu$ are exponents that characterize the disordered system.
It means that disordered systems
at criticality are only weakly self averaging when $\frac{\alpha}{\nu}<0$.
Though our derivation is not valid when $\frac{\alpha}{\nu}=0_{+}$
($C\sim \log l$) it
seems that in this case there is no self averaging of the sample to sample
fluctuations.This point is further discussed in the next section.

 We note that for $\alpha<0$ and in the large $l$ limit considered at the
end of the Appendix [equation (\ref{eq: Qlargel no x}) ],
where \begin{equation}
   Q_{x}( -\Delta t_{c}(x,l)l^{y_{t}} ) \rightarrow
   Q    ( -\Delta t_{c}(x,l)l^{y_{t}} )
\;, \label{eq: QTcNoX} \end{equation}
the coefficients of $Q_{x}$ are independent of $x$ so that $D_{x}=D_{0},
E_{x}=E_{0}$ etc. .
Neglecting the analytic part
of $P$, this limit corresponds to our derivation with only
$D_{0},E_{0},F_{0}\not=0$ and all other coefficients (
$D_{1},D_{2},E_{1}$ etc. ) equal to zero. Thus
in this limit the main result (\ref{eq: Vlead2}) is unchanged, though less
assumptions are needed.


{}From (\ref{eq: Pav}) corrections to the scaling of $P$ are obtained,
 \begin{equation}
  [P_{x}(T_{c},l)]=  a+D_{0}l^{\rho}+
 ( c+ D_{2}l^{\rho}-E_{1}l^{\rho+y_{t}}+F_{0}l^{\rho+2y_{t}} )
  K^{2}_{r}l^{-d}
 \;. \label{eq: Pcorr1} \end{equation}
 So that the leading behaviour of $P$ is
 \begin{equation}
  [P_{x}(T_{c},l)]=  a+D_{0}l^{\rho}+ F_{0}K^{2}_{r}l^{\rho+2y_{t}-d}
 = a+D_{0}l^{\rho}+ F_{0}K^{2}_{r}l^{\rho+\frac{\alpha}{\nu} }
 \;. \label{eq: Pcorr2} \end{equation}
Thus for negative $\alpha$, the third term in (\ref{eq: Pcorr2}) is a
correction to scaling due to sample to sample fluctuations. It follows that
for $\frac{\alpha}{\nu}< 0 $, (\ref{eq: Pcorr2}) and
(\ref{eq: Vlead2}) are consistent with $V/[P]^2 \sim l^{\frac{\alpha}{\nu}
}$. A special case is when $\alpha_{\text{pure}}<0$ and randomness is
an irrelevant operator (at the pure system fixed point) with a scaling
exponent
$(\frac{\alpha}{\nu})_{\text{pure}}$. In this case the disordered system has
the same exponents as the pure one,
$\frac{\alpha}{\nu}=(\frac{\alpha}{\nu})_{\text{pure}}$. Therefore the
correction
to scaling we have obtained due to sample to sample fluctuations has the
same exponent as the correction term connected with the irrelevant operator
corresponding to randomness.

%
\section{Comparison of Theory with Variance Results of the RBAT models}

 The derivation presented in the previous section as can be readily seen
 from equations (\ref{eq:
scaling1}, \ref{eq: Ax}, \ref{eq: Q(X)}), involved an expansion in the two
parameters $\sqrt{[(\Delta t_{c}(x,l))^{2}]}$ and  $\sqrt{[(\Delta
t_{c}(x,l))^{2}]}l^{y_{t}}$. These scale as $K_{r}l^{-\frac{d}{2}}$ and
$K_{r}l^{y_{t}-\frac{d}{2}}=K_{r}l^{\frac{\alpha}{2\nu} }$.
Thus the derivation is valid for small $K_{r}$, meaning small disorder and
small $\frac{\alpha}{\nu}$. For negative $\frac{\alpha}{\nu}$ the
validity of the expansion improves as $l$ increases, while a positive
$\frac{\alpha}{\nu}$ is not possible\cite{Chayes}.


In the case of the random bond Ashkin Teller model, we have asymptotically
$C\approx b \log l$ so that $\frac{\alpha}{\nu}=0_{+}$. It seems that in
this case the expansion is not justified. Practically though,  for the
accessible range of lattice sizes, things depend on the
constant of proportionality $b$. If $b$ is small, then for a finite but large
interval of lattice sizes $l$ the expansion is justified.
Indeed, in the case of the highly random RBAT models  $C_{0\ldots4}$, $b$
falls in the range $  0.138(4)\leq b \leq 0.280(6) $. Upon inspection of
Fig. \ref{fig:CvrCaat}
one may also see that the value of the specific heat of these models shows
very little variation for lattice sizes $l\geq16$. Thus the parameter
$K_r l^{\frac{\alpha}{\nu}}$, which should scale with $l$ as
the specific heat does, increases very slowly with $l$. This implies that
for the accessible range of lattice sizes $l$ our expansion is valid.
The specific heat of
the weakly random model $A_{2}$ effectively diverges with a positive
effective $\frac{\alpha}{\nu}$ but because of its weak degree of randomness
there is good reason to believe that the expansion will be valid due to
a small value of $K_{r}$ expected for models with small randomness.
The $B_{2}$ model with moderate disorder is expected to fall between the
$A_{2}$ model and the $C_{0\ldots4}$ models. Thus there is reason to hope
that our theory is applicable to the variance results in the accessible
range $4\leq l\leq256$. Indeed the agreement we now display between
numerical data and theory is good.

For observables with $\rho>0$ the two leading terms in (\ref{eq: Vav(l)})
are the third and sixth terms. We use hyper-scaling to write
$l^{y_{t}}= l^{ \frac{\alpha}{2\nu} +\frac{d}{2} }$ and substitute in
(\ref{eq: Vlead}) $l^{\frac{\alpha}{\nu} }$  by the behaviour of the
specific heat (\ref{eq:C fss cross}). Thus we propose for the RBAT models
the leading behaviour\cite{term F0}
\begin{equation}
V(T_{c},l)=a_{v} l^{2\rho} \ln
[1+ c_{0}(l^{(\alpha/\nu)_{\text{pure}}}-1)] +   c_{v} l^{2\rho-\frac{d}{2}
} \{ \ln [1+ c_{0}(l^{(\alpha/\nu)_{\text{pure}}}-1)] \}^{ \frac{1}{2} }
\;, \label{eq: var RBAT} \end{equation}
with $a_{v}\equiv E_{0}^{2} K_{r}^{2}$ and $c_{v}\equiv -2D_{1}E_{0}
K_{r}^{2}$
(note that for every thermodynamic quantity there are different coefficients
$E_{0},D_{1}$ etc. )
The expression $\ln [1+ c_{0}(l^{(\alpha/\nu)_{\text{pure}}}-1)]$ describes
the  singular behavior of
the specific heat including the crossover from the pure model behavior ,
characterized by the pure model exponent
$(\alpha/\nu)_{\text{pure}}$ (see Eq. (\ref{eq:C fss cross}) ). $\rho$
is the critical exponent of the quantity whose variance is measured (e.g.
$\rho= \frac{\gamma}{\nu}$ for
$\chi$ and $\rho=\frac{\gamma^{(p)}}{\nu}$ for $\chi^{(p)}$ ).

\begin{figure}[h]
\centering
\epsfysize=3.25truein
\epsfxsize=5.truein
\epsffile{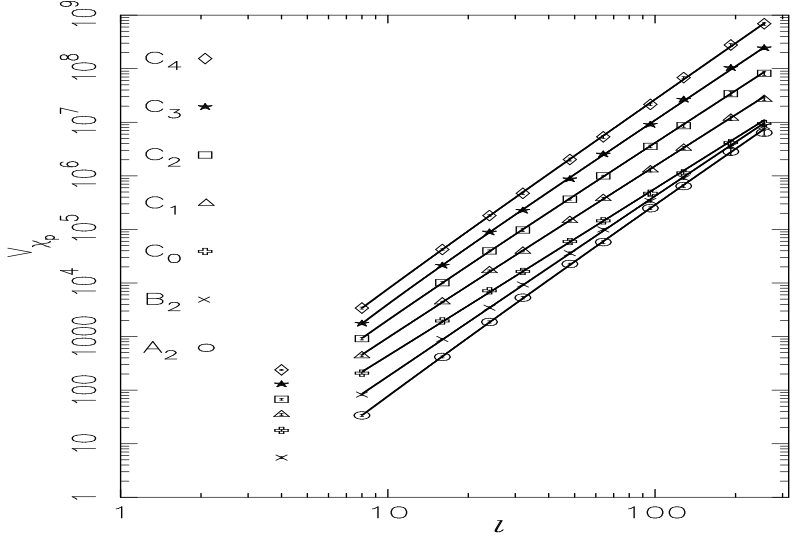}
\caption{ The variance of $\chi^{(p)}$, $V_{\chi^{(p)}}$ as a function of
$\log l$ for all critical
models, $C_{0..4}$, and $A_{2},B_{2}$ of the RBAT model.
For the sake of
clarity $V_{\chi^{(p)}}$ of the model $C_{i}$ was multiplied by $2^{i+1}$.
The solid lines are  fits to the
form ( \protect\ref{eq: var RBAT}) , yielding estimates for fitting
parameters which are listed in table \ \protect\ref{tab: var}
. } \label{fig: XPVfit2} \end{figure}

In fig. \ref{fig: XtVfit2} we show the variance of $\chi$, $V_{\chi}$ of the
seven critical RBAT
models fitted by the function (\ref{eq: var RBAT}), where the parameters
$c_{0}, (\alpha/\nu)_{\text{pure}}$ and $\frac{\gamma}{\nu}$ were taken from
Tables \ref{tab:Cv fit} and \ref{tab: Rand1}. For the
sake of clarity (so that the data do not fall on top of each-other)
$V_{\chi}$ of the model $C_{i}$ was multiplied by $2^{i+1}$. The fitting
parameters $a_{v},c_{v}$ are given in Table \ref{tab: var}. The agreement
with our scaling prediction is quite encouraging.
\vbox{
 \begin {table}
\caption{  Fitting parameters for the variances of $\chi, \chi^{(p)}, E$
for the critical models $C_{0..4}$, and $A_{2},B_{2}$ according to eq-s.
\ (\protect\ref{eq: var RBAT} ) and (\protect\ref{eq: varE RBAT} ),
using lattice sizes $l\geq8$. }
 \label{tab: var}
\begin{tabular}{lllllll}
  &\multicolumn{2}{c}{$V_{\chi}$} &\multicolumn{2}{c}{$V_{\chi^{(p)}}$}
&\multicolumn{2}{c}{$V_{E}$} \\
       & $a_{v}$& $c_{v}$ &  $a_{v}$& $c_{v}$ & $a_{v}$& $b_{v}$\\ \hline
$C_{0}$(Ising)&0.0145(7)&0.11(2)  & 0.033(1) &.07(4)  &0.29(7)  &0.8(2)\\
$C_{1}$       &0.0039(1)&0.026(10)& 0.0134(3)&-0.03(2)&0.128(14)&-1.45(19)\\
$C_{2}$       &0.0059(1)&0.014(10)& 0.0172(2)&-0.09(2)&0.135(13)&-1.19(16)\\
$C_{3}$       &0.0069(2)&-0.01(1) & 0.0147(3)&-0.12(2)&0.133(15)&-1.27(18)\\
$C_{4}$(Potts)&0.0082(2)&-0.04(1) & 0.082(2) &-0.04(1)&0.121(13)&-1.25(17)\\
$B_{2}$       &0.033(1) &0.13(2)  & 0.092(2) & 0.19(4)&0.99(15) & 1.2(2)\\
$A_{2}$       &0.056(1) &0.028(12)& 0.143(3) & 0.05(3)&5.49(44) &1.28(23)
 \end{tabular}
\end{table}
 }
The same analysis has been carried out for the variance of $\chi^{(p)}$,
 where $\frac{\gamma^{(p)}}{\nu}$ was taken from Table \ref{tab: Rand1},
and the results are plotted in fig. \ref{fig: XPVfit2}. Again the fitting
parameters $a_{v},c_{v}$ are given in Table \ref{tab: var} and the agreement
 between the numerical results and our scaling prediction is encouraging.
 We stress that the only fitting parameters of the fits in figures
 \ref{fig: XtVfit2} and \ref{fig: XPVfit2} are $a_{v}$ $c_{v}$; the other
 parameters of eq. (\ref{eq: var RBAT}), $\rho , c_0 , (\alpha/\nu)_p$
 were obtained previously\cite{RBAT} from the specific heat results and from
 the results for $\chi$ and $\chi^{(p)}$.

\begin{figure}
\centering
\epsfysize=3.25truein
\epsfxsize=5.truein
\epsffile{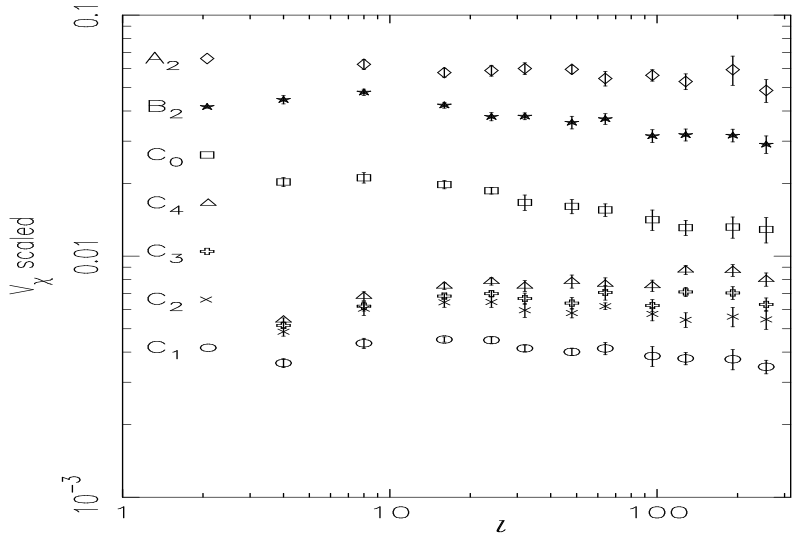}
\caption{ The scaled variance of $\chi$, $V_{\chi}
l^{-2\frac{\gamma}{\nu}}/\ln [1+c_{0}(l^{(\alpha/\nu)_{\text{p}}}-1)]$  as a
function of $\log l$ for all critical
models, $C_{0..4}$, and $A_{2},B_{2}$ of the RBAT model.
 } \label{fig: XtVSca1} \end{figure}

 Since the first term in (\ref{eq: var RBAT}) is the dominant one (by
a factor of $l^{y_{t}}$, where $y_{t}\geq 1$ ), we test (\ref{eq: var
RBAT}) again in another  manner. in
fig.  \ref{fig: XtVSca1} we plot the scaled $V_{\chi}$ : $V_{\chi}
l^{-2\rho}/\ln
[1+c_{0}(l^{(\alpha/\nu)_{\text{pure}}}-1)]$. Indeed it seems that the data
points approach a constant value, confirming the leading term in (\ref{eq:
var RBAT}) which originates in the leading behavior of the variance
(\ref{eq: Vlead2}).

For the energy $\rho=(\alpha-1)/\nu<0$ so that the two leading terms in
(\ref{eq: Vav(l)}) are the third and the fifth ones.
Again by using hyper-scaling and
substituting $l^{\frac{\alpha}{\nu} }$  by the behaviour of the
specific heat (\ref{eq:C fss cross}), we arrive at the scaling form for the
variance of the energy
\begin{equation}
V_{E}(T_{c},l)= \{ a_{v}(\ln
[1+c_{0}(l^{(\alpha/\nu)_{\text{pure}}}-1)])^{2}
  + b_{v}\ln [1+c_{0}(l^{(\alpha/\nu)_{\text{pure}}}-1)] \} l^{-d}
\;, \label{eq: varE RBAT} \end{equation}
with $b_{v}=-2bE_{0} K_{r}^{2}$
In fig. \ref{fig: EnRVft2} we show the variance of the energy, $V_{E}$ of
the seven critical RBAT
models fitted by the function (\ref{eq: varE RBAT}). For the sake of clarity
$V_{E}$ of the model $C_{i}$ was multiplied by $2^{i+1}$. The agreement
between theory and the numerical data is good and the fitting parameters
$a_{v},b_{v}$ are given in table \ref{tab: var}.

\begin{figure}
\centering
\epsfysize=3.25truein
\epsfxsize=5.truein
\epsffile{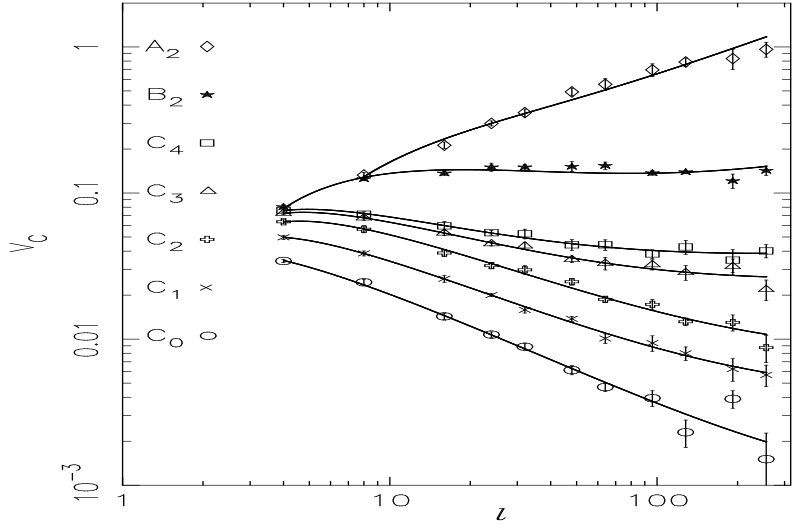}
\caption{ The variance of the specific heat, $V_{C}$ as a function of $\log
l$ for all critical models, $C_{0..4}$, and $A_{2},B_{2}$ of the RBAT model.
The solid lines are fits to
the form ( \protect\ref{eq: varC RBAT}) , yielding estimates for the fitting
parameters $a_{v},b_{v},c_{v}$ which are listed in table \ \protect\ref{tab:
varCM} . } \label{fig: CvVftt3} \end{figure}

For the magnetization and the specific heat $\rho=\frac{\alpha}{\nu}$ and
$\rho=\frac{-\beta}{\nu}$ respectively. In these cases $|\rho|$ is small
and the fifth and the sixth terms in Eq.
(\ref{eq: Vav(l)}) are of similar order in $l$ . Thus one may not neglect
one term with respect to the other as was done for the energy and the
susceptibility. Thus we fit the variance of the specific heat to the form
\begin{eqnarray}
V_{C}(T_{c},l)= a_{v}\{ \ln
[1+c_{0}(l^{(\alpha/\nu)_{\text{pure}}}-1)]\}^{3}
  + b_{v} l^{-\frac{d}{2}} \{ \ln
[1+c_{0}(l^{(\alpha/\nu)_{\text{pure}}}-1)]\}
^{1.5} +                  \nonumber \\
   c_{v} l^{-\frac{d}{2}} \{ \ln
[1+c_{0}(l^{(\alpha/\nu)_{\text{pure}}}-1)]\} ^{2.5}
 \;. \label{eq: varC RBAT} \end{eqnarray}
In fig. \ref{fig: CvVftt3} the variance of the specific heat,
$V_{C}$ of the seven critical RBAT models is fitted by the function
(\ref{eq: varC RBAT}), with the fitting coefficients given in Table
\ref{tab: varCM}.
 \vbox{
 \begin {table}
\caption{  Fitting parameters for the variances of the specific heat $V_{C}$
and  the magnetization $V_{M}$
for the critical models $C_{0..4}$, and $A_{2},B_{2}$ according to eq-s.
\ (\protect\ref{eq: varC RBAT} ) and (\protect\ref{eq: varM RBAT} ),
using lattice sizes $l\geq4$. }
 \label{tab: varCM}
\begin{tabular}{lllllll}
  &\multicolumn{3}{c}{$V_{C}$} & \multicolumn{3}{c}{$V_{M}$}  \\
       & $a_{v}$& $b_{v}$ &  $c_{v}$& $a_{v}$ & $b_{v}$& $c_{v}$\\ \hline
$C_{0}$(Ising)&0.000016(12)& 0.028(10)&0.008(4) &0.0037(1) &0.25(2)   &-0.28(2)
 \\
$C_{1}$       &0.0000011(1)&-0.045(4) &0.0038(3)&0.00108(3)&0.102(9)  &-0.12(1)
 \\
$C_{2}$       &0.0000028(5)&-0.078(8) &0.0081(7)&0.00169(7)&0.10(2)   &-0.12(2)
 \\
$C_{3}$       &0.0000079(6)&-0.069(5) &0.0069(5)&0.00213(9)&0.054(20)
&-0.068(24)\\
$C_{4}$(Potts)&0.0000099(4)&-0.062(5) &0.0057(4)&0.0027(1)
&-0.004(23)&-0.004(28)\\
$B_{2}$       &0.0079(6)   &-0.44(9)  &1.4(1)   &0.0089(3) &0.27(3)   &-0.30(4)
 \\
$A_{2}$       &0.67(7)     &-4.9(18)  &36.5(54) &0.0167(8) &0.07(7)
&-0.07(10)
 \end{tabular}
\end{table}
 }
The data for large lattice sizes is rather noisy and three parameter
fits are not so reliable with only eleven data points, so that both the
results and the fitting curves in Fig. \ref{fig: CvVftt3} should be taken
with
a grain of salt. The obtained fitting coefficients are consistent with the
coefficient
$E_{0}$ being much smaller than $b,D_{1}$. A small value for $E_{0}$ is
quite plausible if
the specific heat as a function of the temperature is close to
being symmetric around the critical point [see (\ref{eq: Q(X)})]. This
symmetry is supported by the symmetric form of the histograms of the
specific heat as shown in figures \ref{fig: HistIsCv} and \ref{fig:
HistPoCv}.
For the Ising model $C_{0}$ the errors of the coefficients $a_{v},b_{v}$ and
$c_{v}$ are of the same order of magnitude as the coefficients themselves.
However for the other models the errors are reasonable and though $a_{v}$ is
small, we
have $E_{0}^{2} K_{r}^{2} \equiv a_{v} > 0 $, meaning that, asymptotically,
the first term in (\ref{eq: varC RBAT}) will dominate. This implies that
the specific heat of the RBAT models is non self averaging, excluding
possibly the random bond Ising model.
possibly, the theory needs some changes in order to be applied
to the specific heat $C$ which diverges logarithmically (and as a double
logarithm for the random bond Ising model) and not
with a simple power law.

In fig. \ref{fig: MRVft3n} the variance of the magnetization,
$V_{M}$ of the seven critical RBAT models is fitted by the function
\begin{eqnarray}
V_{M}=a_{v} l^{-2\frac{\beta}{\nu}} \ln
[1+ c_{0}(l^{(\alpha/\nu)_{\text{pure}}}-1)]+
 b_{v} l^{ -\frac{\beta}{\nu}-\frac{d}{2} }
 \{ \ln [1+ c_{0}(l^{(\alpha/\nu)_{\text{pure}}}-1)] \}^{ \frac{1}{2} }+
				      \nonumber \\
 c_{v} l^{-2\frac{\beta}{\nu}-\frac{d}{2} }
 \{ \ln [1+ c_{0}(l^{(\alpha/\nu)_{\text{pure}}}-1)] \}^{ \frac{1}{2} }
\;. \label{eq: varM RBAT} \end{eqnarray}
The fitting coefficients $a_{v},b_{v}$ and $c_{v}$ are given in Table
\ref{tab: varCM}. The data are much more noisy than the data of the
susceptibility (see Fig. \ref{fig: XtVfit2}).

\vbox{
\begin{figure}
\centering
\epsfysize=3.25truein
\epsfxsize=5.truein
\epsffile{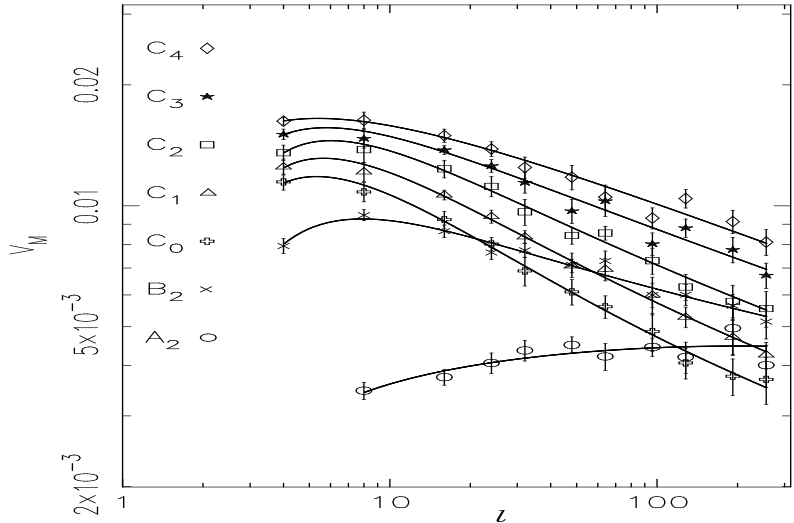}
\caption{ The variance of the magnetization, $V_{M}$ as a function of $\log
l$ for all critical models, $C_{0..4}$, and $A_{2},B_{2}$ of the RBAT model.
The solid lines are fits to
the form ( \protect\ref{eq: varM RBAT}) , yielding estimates for the
fitting
parameters $a_{v},b_{v},c_{v}$ which are listed in table \ \protect\ref{tab:
varCM} . } \label{fig: MRVft3n} \end{figure}
      }

 To summarize, we have examined the sample to sample fluctuations in various
thermodynamic quantities of some random bond Ashkin Teller models.
These include the random bond Ising and four state Potts models.
It was found that far from criticality all thermodynamic quantities examined
are strongly self averaging (that is their variance scales as $l^{-d}$) .
At the critical point we found that the susceptibility $\chi$, the
susceptibility of the polarization $\chi^{(p)}$ and the magnetization $M$
are non self averaging, while the energy $E$ is weakly self averaging.
The data for the variance of the specific heat seems
to imply weak self averaging of the specific heat. Since the data are not
accurate at the larger sizes used, this may well
be a transient behavior, compatible with our theory which predicts
that asymptotically the specific heat is non self averaging.
A phenomenological finite size scaling theory was developed for the sample
to sample fluctuations. Its main prediction is that when the
specific heat exponent $\alpha<0$ ($\alpha$ of the disordered model) then,
for a quantity $P$ which scales as $l^{\rho}$ at criticality, its
variance $V_{P}$ will scale asymptotically as
$l^{2\rho+\frac{\alpha}{\nu}}$. The theory is not applicable in
the asymptotic limit ($l\rightarrow \infty$) to cases where
$\frac{\alpha}{\nu}=0_{+}$. Nonetheless in the accessible range of lattice
sizes we found very good agreement between the theory and the data for
$V_{\chi}, V_{\chi^{(p)}}$ and $V_{E}$. The data for $V_{\chi}$ is especially
convincing. The theory also describes well the variance of models with weak
disorder, exhibiting slow crossover to the randomness dominated behavior.
The theory may also be compatible with the data for $V_{M}$
and $V_{C}$, but evidence for this is less convincing. We note that if our
assumption (\ref{eq: width}) is incorrect and should be replaced
asymptotically by\cite{private} $\delta T_{c}(l)\sim l^{-y_{t}} $, then our
theory predicts that $V_{P}\sim l^{2\rho}$ independent of $\alpha$. In this
case all quantities (excluding the energy which has a non vanishing
non singular part) are non self averaging independent of $\alpha$. \\
In order to further
test our theory we intend to study the sample to sample fluctuations in
the site dilute three dimensional Ising model where $\frac{\alpha}{\nu}<0$
and our analysis holds.
\acknowledgments
We thank A.B. Harris and A. Aharony for a most helpful discussion,
and D. Stauffer for critical reading of the manuscript at its early stages.
SW would like to thank H.J. Herrmann and the Many Particle Group of the HLRZ
J\"{u}lich for their warm hospitality and the generous allocation of
computer time on the Intel IPSC/860 and the Paragon.
This research has been supported by the US-Israel Bi-national Science
Foundation (BSF), and the Germany-Israel Science Foundation (GIF).

\appendix
\section*{}
 \label{app:qualitative}
 In this section we draw some more conclusions based on the finite size
scaling form (\ref{eq: Psing}), without making any assumptions on the
explicit $x$ dependence of the coefficients of $\tilde{Q}_{x}$.

 What can we deduce about the coefficients of $\tilde{Q}_{x}$ from the Brout
argument?  Consider (\ref{eq: Psing}) in the limit
$l\rightarrow\infty$, $\dot{t}_{x}$ finite. In this
case $\xi$ is finite, the Brout argument holds, and one expects $P_{x}(T)$
to be sample independent. This means
that in this limit we expect the coefficients of $\tilde{Q}_{x}$ and its
argument to converge
to some $x$ independent values. It follows that we can assume that these
coefficients are distributed according to some unknown distribution function
whose width $w(l)$ depends on $l$ and tends to zero as $l\rightarrow\infty$.

 Is there any limit in which one may recover the usual finite size scaling
behaviour, completely independent of the specific sample $x$?
Consider the limit $l$ large but finite and $T\approx
T_{c}$. Let us now add the assumption that the width $w(l)$ tends to zero no
slower than $l^{-d/2}$ . Then for large enough $l$, according to equations
(\ref{eq: width}-\ref{eq: tx dot}), $\dot{t}_{x}$ approaches $\dot{t}$
given by \begin{equation}
 \dot{t}= \frac{T-T_{c}(l)}{T_{c}}
 \label{eq: t dot} \end{equation}
as $l^{-d/2}$, and
\begin{equation}
  \tilde{Q}_{x}( \dot{t}_{x}l^{y_{t}} ) \rightarrow \tilde{Q}( \dot{t}
l^{y_{t}} )
\;, \label{eq: Qlargel} \end{equation}
 so that we recover the usual\cite{usual} finite size scaling behaviour.

 The limit (\ref{eq: Qlargel}) cannot account for the large
sample to sample fluctuations that we have numerically observed at $T=T_{c}$
even for rather large values of $l$.
Indeed, special care is needed in the case $T=T_{c}$, where $\xi/l$ is not
small and the Brout argument does not hold. It turns out that in this case
the limit, where the $x$
dependence of $\dot{t}_{x}$ can be ignored as in (\ref{eq: Qlargel}), does
not occur or is reached `slowly'. When $T=T_{c}$, then
$\dot{t}_{x}=-t_{c}(x,l)=\{t_{c}(l)-t_{c}(x,l)\}-t_{c}(l)$,
so that according to (\ref{eq: width}) and (\ref{eq: Tc scaling})
$\dot{t}_{x}$ is a difference of a fluctuating term of order
$l^{-d/2}$ and a term of a constant sign of order $l^{-y_{t}}$. Therefore
   \begin{equation}
\frac{|\delta \dot{t}_{x}| }{ |\dot{t}_{x}| }\sim \left\{ \begin{array}{ll}
 \,l^{y_{t}-d/2}=l^{\frac{\alpha}{2\nu}}= l^{-|\frac{\alpha}{2\nu}|}
			   &  \mbox{if }\;\; d/2>y_{t} \;\;\;(\alpha<0) \\
 \,1 & \mbox{if } \;\; d/2\leq y_{t} \;\;\;(\alpha\geq 0)\end{array}
\right. \label{eq: DeltaT} \;.   \end{equation}

  Thus for $T=T_{c}$ and positive $\alpha$, \ $\dot{t}_{x} \not\rightarrow
\dot{t}$ for large $l$ and (\ref{eq: Qlargel}) is not justified.
In this case the relative fluctuations in the argument of $\tilde{Q}_{x}$
are of order 1 and their absolute magnitude scales as
$l^{\frac{\alpha}{2\nu}}$ so that it increases with $l$. So for large $l$
the argument of $\tilde{Q}_{x}$ is a constant plus a large fluctuating
quantity which increases with $l$. It follows that
$\tilde{Q}_{x}$ cannot be expanded as is done in Sec. \ref{sec: scale var},
and that the limit (\ref{eq: Qlargel}) does not exist.

For negative $\alpha$ it follows that the fluctuations in the argument of
$\tilde{Q}_{x}$ scale as $l^{-|\frac{\alpha}{2\nu}|}$.
Since we have assumed that the fluctuations in
the coefficients of $\tilde{Q}_{x}$, $w(l)$, scale as $l^{-d/2}$ then if
$|\frac{\alpha}{2\nu}| <\frac{d}{2}$ then at $T=T_{c}$ \ $w(l)$ decreases
faster than the fluctuations in the argument of $\tilde{Q}_{x}$. Thus one
may consider the range of $l$ for which
\begin{equation}
  \tilde{Q}_{x}( \dot{t}_{x}l^{y_{t}} ) \rightarrow \tilde{Q}( \dot{t}_{x}
l^{y_{t}} )
\;, \label{eq: Qlargel no x} \end{equation}
where only the argument of $\tilde{Q}_{x}$ is $x$ dependent. In this case
the coefficients of $\tilde{Q}$ are some constants for which we
need not assume anything about their $x$ or $l$ dependence. Consideration of
the limit (\ref{eq: Qlargel no x}) suffices to reach our main result
(\ref{eq: Vlead}) (but not corrections to (\ref{eq: Vlead})), independent of
the assumptions made
in Sec. \ref{sec: scale var}  on the $x$ dependence of the coefficients of
$\tilde{Q}_{x}$.


\begin{thebibliography}{99}
\bibitem{pure} Since the anisotropic model is defined by two sets of
coupling constants, as is the random-bond model, it serves as a convenient
reference pure model. It is a pure model in the sense that anisotropy is an
irrelevant operator. Therefore it is of the same universality class as the
homogeneous (or pure) model.
\bibitem{magnetization} As usual in Monte Carlo simulations, the
magnetization $M$ is actually defined as the average of the absolute value
of the magnetization, $M=\frac{1}{l^d} \av{| \sum_i \sigma_i |}$.
\bibitem{practical}  These considerations have a large practical
importance concerning the usual way of calculating the susceptibility in MC
simulations. It is customary to define the susceptibility (per spin) as
$\chi\equiv l^{d}\av{M^{2}}$, which is well justified\cite{BH:book} when
$T_{c}$ is slightly higher than the temperature of the maximum of the true
susceptibility of the sample. This definition cannot be used in the low
temperature phase where one needs to subtract $\av{|M|}^{2}$. But in the
case of large fluctuations
in $T_{c}(x,l)$,  there exist samples in which $T_{c}$ is smaller than the
temperature of the maximum of the true susceptibility, and
therefore $l^{d}\av{M^{2}}> \chi=l^{d}(\av{M^{2}}-\av{|M|}^{2})$.
This effect is probably the reason for the asymmetry of the histograms of
the susceptibility ( figures \ref{fig: HistIsXi}-\ref{fig: HistPoXi24}).
Those samples in which $T_{c}< T_{c}(x,l)$ are effectively in their low
temperature phase and give an estimate of $\chi$ which is biased to higher
values than the true $\chi$.
\bibitem{E not nor} We do not consider $V_{E}/ [E]^2$ since $[E]\rightarrow
{\text const}$ as $l \rightarrow \infty$.
\bibitem{usual}
 In the more commonly used form of finite size scaling the argument of
$\tilde{Q}$ would be $t_{x}=\frac{T-T_{c}(x)}{T_{c}}$. But in the case of a
disordered sample  it is
unclear how to define a critical temperature which depends on the sample $x$
$T_{c}(x)$ but does not depend on its size $l$. Hence we use scaling with
respect to the reduced temperature $\dot{t}_{x}=\frac{T-T_{c}(x,l)}{T_{c}}$,
where $\dot{t}_{x}$ is defined using the pseudo-critical temperature
$T_{c}(x,l)$. In any case this version of finite size scaling is equivalent
to the more common one\cite{Barber}.
\bibitem{analytic} We assume the analytic part of $P_{x}$ depends on
$(t-\Delta t_{c}(x,l))$
and not on $\dot{t}_{x}$. If we assume that $P_{x}$ has an analytic
dependence on $\dot{t}_{x}$ we find that $[P^{\text{analytic}}]$ has terms
that scale as $l^{-y_{t}}$. Such dependence of the analytic part on
exponents connected with critical behavior does not seem plausible.
\bibitem{Tc(x)} Note that from the structure of (\ref{eq: scaling1}) it is
clear that $\Delta t_{c}(x,l)$ plays the role of a `critical temperature of
the sample $x$' .
\bibitem{term F0} For $\frac{\alpha}{\nu}\geq 0$ it seems that one would
need to include also the last term in (\ref{eq: Vav(l)}) proportional to
$l^{2\rho+2 \frac{\alpha}{\nu}}$. But since this term is also proportional
to $K_{r}^{4}$, this may not be necessary for small values of
$\frac{\alpha}{\nu}$ or for small values of $K_{r}$. indeed, attempts to
include such a term in our fits were not successful, yielding a negative
coefficient where only a positive one is consistent with the theory.
\end{thebibliography}
\end{document}